\documentclass[journal]{vgtc}
\usepackage[utf8]{inputenc}
\usepackage{amsmath}
\usepackage{amsfonts}
\usepackage{graphicx}
\usepackage{hyperref}
\usepackage{subcaption}
\usepackage{algpseudocode}
\usepackage{algorithm}
\usepackage{algpseudocode}

\vgtccategory{Research \\ Type: Technique \& Algorithm}
\onlineid{1382}
\usepackage{epstopdf}
\hypersetup{
    colorlinks=true,
    linkcolor=blue,
    filecolor=magenta,      
    urlcolor=cyan,
}

\usepackage[colorinlistoftodos]{todonotes}

\DeclareMathOperator{\cD}{\mathcal{D}}
\DeclareMathOperator{\cP}{\mathcal{P}}
\DeclareMathOperator{\cO}{\mathcal{O}}

\DeclareMathOperator{\mD}{\mathbf{D}}
\DeclareMathOperator{\mP}{\mathbf{P}}
\DeclareMathOperator{\mQ}{\mathbf{Q}}
\DeclareMathOperator{\mX}{\mathbf{X}}
\DeclareMathOperator{\mY}{\mathbf{Y}}

\title{Multi-Perspective, Simultaneous Embedding 
}
\author{Md Iqbal Hossain, Vahan Huroyan, Stephen Kobourov, Raymundo Navarrete}
\authorfooter{
\item
 Md Iqbal Hossain and Stephen Kobourov are with the Computer Science Department of The University of Arizona, E-mails:  hossain@email.arizona.edu and kobourov@cs.arizona.edu.
\item
  Vahan Huroyan and Raymundo Navarrete are with the Department of Mathematics of The University of Arizona, E-mails: vahanhuroyan@math.arizona.edu and raymundo@math.arizona.edu.
}
\date{July 2020}

\abstract{
We describe MPSE: a Multi-Perspective Simultaneous Embedding method for visualizing high-dimensional data, based on multiple pairwise distances between the data points. Specifically, MPSE computes positions for the points in 3D and provides different views into the data by means of 2D projections (planes) that preserve each of the given distance matrices. We consider two versions of the problem: fixed projections and variable projections. MPSE with fixed projections takes as input a set of pairwise distance matrices defined on the data points, along with the same number of projections and embeds the points in 3D so that the pairwise distances are preserved in the given projections. MPSE with variable projections takes as input a set of pairwise distance matrices and embeds the points in 3D while also computing the appropriate projections that preserve the pairwise distances. The proposed approach can be useful in multiple scenarios: from creating simultaneous embedding of multiple graphs on the same set of vertices, to reconstructing a 3D object from multiple 2D snapshots, to analyzing data from multiple points of view. We provide a functional prototype of MPSE that is based on an adaptive and stochastic generalization of multi-dimensional scaling to multiple distances and multiple variable projections. We provide an extensive quantitative evaluation with datasets of different sizes and using different number of projections, as well as several examples that  illustrate the quality of the resulting solutions.
}

\keywords{Graph visualization, Dimensionality reduction, Multidimensional scaling, Mental map preservation.}

\teaser{
  \centering
    \includegraphics[width=.19\linewidth]{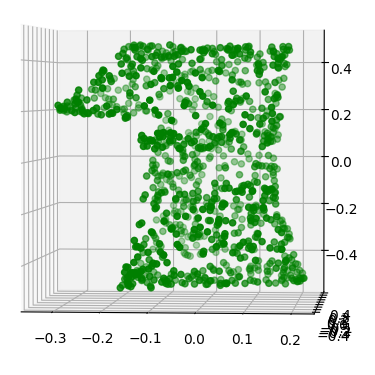}
    \includegraphics[width=.19\linewidth]{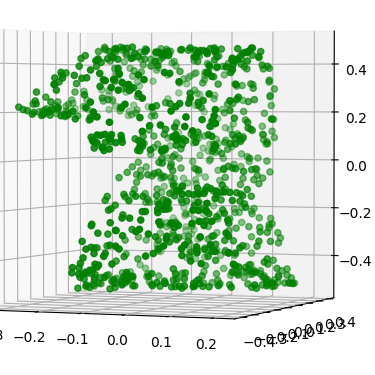}
    \includegraphics[width=.19\linewidth]{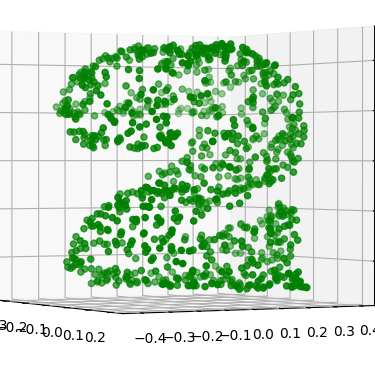}
    \includegraphics[width=.19\linewidth]{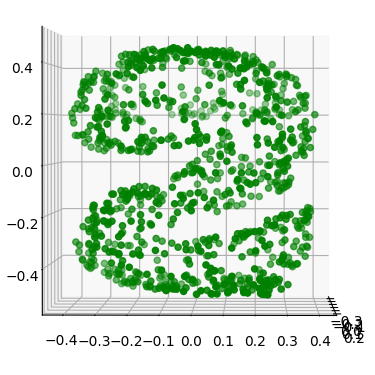}
    \includegraphics[width=.19\linewidth]{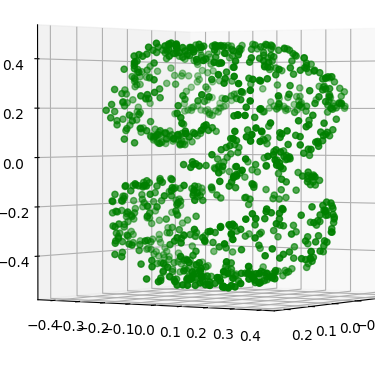}
          
  \caption{Given several distances between a set of objects and matching projection planes, MPSE computes 3D coordinates for all the objects, such that when projected to the corresponding 2D planes the distances in each plane match the corresponding input distances. The input dataset is inspired by a sculpture ``1, 2, 3" by James Hopkins.}
  \label{fig:teaser}
}

\begin{document}

\firstsection{Introduction}

\maketitle

Given a high dimensional dataset, one of the main visualization goals is to produce an embedding in 2D or 3D Euclidean space, that suitably captures pairwise relationships among the represented data. Dating back to the 1960s, a classical tool that is widely used for both graphs and high dimensional dataset visualization is Multidimensional Scaling (MDS), which aims to preserve the distances between all pairs of datapoints or nodes/objects~\cite{shepard1962analysis}. 
Dimensionality reduction is a more general version of this problem, aiming to project a given dataset to a lower dimensional space. 
There are many popular algorithms for dimensionality reduction, from linear methods such as  principal component analysis (PCA)~\cite{jolliffe1986principal}, to non-linear methods such as t-SNE~\cite{maaten2008visualizing} which captures local neighborhoods and clusters and  UMAP~\cite{mcionnes2018umap} which aims to capture both local and global structure.

\begin{figure*}
    \includegraphics[width=0.24\linewidth]{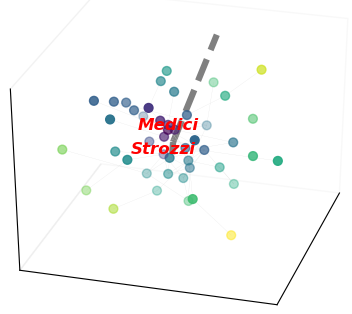}
    \includegraphics[width=0.24\linewidth]{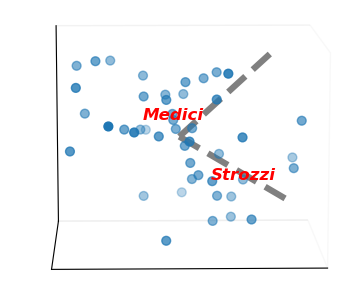}
    \includegraphics[width=0.24\linewidth]{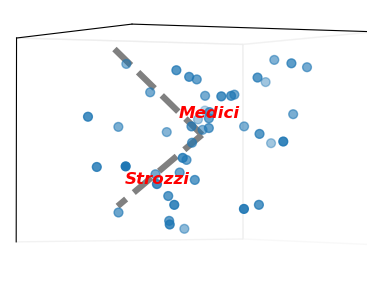}
    \includegraphics[width=0.24\linewidth]{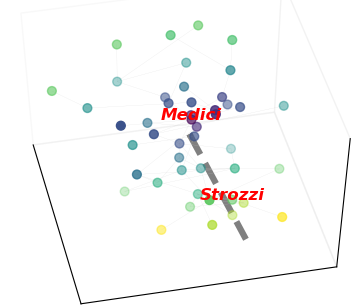}
\caption{ visualization of marriage and loan relations among prominent families in 15th-century Florence, viewed from different angles. Each node represents one of the families. An edge is drawn between two families if at least one corresponding relation existed between the families. 
The leftmost image shows the view that best captures pairwise distances of the marriage bonds between the families. The rightmost image shows the view that best captures pairwise distances of the loan bonds between the families. The middle two images show other views of the 3D visualization. The embedding is computed using MPSE with variable projections. A discussion about the quality of the embedding can be found in Section \ref{sec:florence}.}
\label{fig:florence_3d}
\end{figure*}

We consider situations where instead of just one pairwise distance function, the input is a set of pairwise distance functions on the same set of objects. The optimization goal is to place the points in 3D so that the embedding simultaneously preserves the distances between the objects when projected to some planes. 
For example, Fig.~\ref{fig:florence_3d} shows different views of a 3D visualization of a network dataset with multiple attributes. In this visualization, when viewing the 3D coordinates from the correct perspective, the apparent distance between nodes represents the true network distances for one of the attributes. See section \ref{sec:florence} for details about the dataset and the visualization.

For the case of graph visualisation, consider a set of vertices $V$ (e.g., researchers in one university) and several relationships defined between them $E_1$, $E_2$, $E_3$  (e.g. joint research publications, joint research proposals, membership in different departments). We would like to compute a layout $L$ in 3D as well as 3 planes ($P_1$, $P_2$ and $P_3$) such that when $V$ is projected onto plane $P_1$ we see the graph $G=(V, E_1)$ so that distances between vertices in the plane $P_1$ correspond to the distances defined by $E_1$. Similarly, when $L$ is projected onto plane $P_2$ we see the graph $G=(V, E_2)$ so that distances between vertices in the plane $P_2$ correspond to the distances defined by $E_2$, and the same for $P_3$ and $E_3$.

In both settings, this is a strict generalization of the classical MDS problem, which can be seen as a special case with only one pairwise distance function. Even for this special case this problem is known to be difficult as the standard optimization approaches such as gradient descent are not guaranteed to converge to the global optimum. Nevertheless, in practice, when there is a clear structure in the given graph, MDS is often likely to find a good local optimum and as we show in this paper, the simultaneous optimization of our Multi-Perspective, Simultaneous Embedding (MPSE) produces good solutions.

A common approach for visualizing different relationships on the same set of objects involves small multiples and often some mechanism (such as brushing and linking) to connect the same objects in the different views, or morphing from one view to the other. In contrast, MPSE produces one 3D layout and each of the different views is a 2D projection. In this way, MPSE attempts to balance the two main desirable qualities of good visualization of multiple relationships defined on the same set of data: the {\em readability} of each individual view (typically captured by a faithful embedding in 2D) and {\em mental map preservation} (typically captured by keeping the objects in the same position across different views).
This cannot be accomplished effectively in 2D as there simply is not enough space to realize more than one relationship well, while it becomes plausible in 3D, as our experiments indicate. MPSE can also allow one embedding to tell multiple stories, depending on the different relationships in the data, captured by the corresponding projections, as shown in Figures~\ref{fig:florence_3d}, \ref{fig:ccapp} and \ref{fig:florence_comparison}. Finally, MPSE can be used for reconstructing 3D structure from a collection of 2D images, as shown in Figures~\ref{fig:teaser}, \ref{fig:123_varying_results} and  \ref{fig:circle_squire_all_results}.

With the advent of virtual reality and augmented reality systems, 3D visualization and 3D interaction with data in general and graphs in particular are becoming a reality. Still, when presenting 3D results in a paper we are limited to showing 2D snapshots. The MPSE webpage \url{http://mpse.arl.arizona.edu} provides 3D visualizations with interactive examples, and a functional MPSE prototype.

We describe the MPSE method in detail. We consider two different settings: one where the projection planes are given as part of the input (e.g., the three sides of a 3D cube) and the second where computing the projection planes is part of the optimization. Both settings have been implemented and work well in practice. A fully functional prototype is available on the webpage and source code is available on github. We provide a description on the python implementation and provide  several examples of MPSE embeddings with real-world examples. Finally, we provide quantitative evaluation of the scalability of the two variants with increasing  number of data points and increasing number of projections. 



\subsection{Previous work}
\label{sec:previous_work}

We review work on visualizing multivariate and multilayer networks, network layout algorithms, multidimensional scaling, simultaneous embedding, and 3D reconstruction algorithms.

\smallskip\noindent{\bf Multivariate network visualization.} Multivariate~\cite{kerren2014introduction} and multilayer~\cite{ghoniem2019state} graph visualization has received a great deal of attention in the last couple of decades. Multi‐label, multi‐edge, multi‐relational, multiplex, multi‐modal and many other variants are cleverly encapsulated by the general multilayer network definition of Kivel{\"a} et al.~\cite{kivela2014multilayer}. 

Wattenberg’s PivotGraph~\cite{wattenberg2006visual} system can visualize and analyze multivariate graphs not using a global graph layout but rather a grid-based approach focusing on different relationships between node attributes.
Semantic substrates~\cite{shneiderman2006network} 
unfold multiple attributes of a graph, a pair of attributes at a time, using two dimensions.
Pretorius and van Wijk~\cite{pretorius2008visual} describe an interactive system that relies on clustering of both nodes and edges and interactive exploration using  brushing and linking (as well as parallel histograms) to show different graph attributes. \emph{GraphDice}~\cite{bezerianos2010graphdice} is an interactive multivariate graph visualization system that allows 
the selection of attributes from an overview plot matrix. This results in a cross dimensional node-link plot for every combination of attributes arranged as a matrix. 
This system is built on the earlier \emph{ScatterDice} system~\cite{elmqvist2008rolling}, which provides the ability to interactively explore multidimensional data using scatterplots. Transitions between the scatterplots with one common component are performed as animated rotations in 3D space.

Different from our approach, most of the earlier methods focus on interactive visualizations of multivariate graphs where changing queries result in changing layouts and views. The idea behind our MPSE approach is to produce one 3D layout of the input graph and several projection planes, such that each attribute corresponds to a projection plane in which geometric distances correspond to the graph distances specified by the particular attribute. The main advantage of this approach is that it  helps preserve the viewer's 3D mental map, while also capturing different relationships in different projections of the same underlying layout.

\smallskip\noindent{\bf Network layout algorithms.} Most basic network layouts are obtained using force-directed algorithms. Also known as spring embedders,
such algorithms calculate the layout of the underlying graph using only information contained within
the structure of the graph itself, rather than relying on domain-specific knowledge~\cite{k-fd13}. Visual analytics systems for graphs usually focus on interaction~\cite{von2011visual}. MDS-like approaches to drawing graphs are exemplified in algorithms such as that of Kamada-Kawai~\cite{kk-adgug-89} and Koren and Carmel~\cite{koren2003visualization}. Most commonly used graph drawing systems, such as  Graphviz~\cite{graphviz}, Pajek~\cite{batagelj1998pajek}, Tulip~\cite{auber2017tulip} and Gephi~\cite{bastian2009gephi}, provide options to visualize graphs in 3D based on MDS-like optimization. Variants of MDS are used in many graph layout systems, including ~\cite{chen2009local,gansner2004graph,pich2009applications,wang2018revisiting}. Other approaches to exploring layouts in 3D include 3D hyperbolic and spherical spaces~\cite{cruz19953d,kobourov2005non,munzner1998exploring}. None of these earlier approaches however,  provides a way to simultaneously optimize different views for the same set of nodes.

\smallskip\noindent{\bf Simultaneous embedding.} This problem is also related to simultaneous graph embedding and matched drawings of graphs~\cite{bkr13}. Specifically, in simultaneous geometric embedding of two or more planar graphs requires planar straight-line drawings of each of the graphs, such that common vertices have the same 2D coordinates in all drawings. This setting is very restrictive and solutions are guaranteed to exist for very restricted type of input graphs, such as two paths~\cite{brass2007simultaneous}, while instances with no solutions can be constructed from a pair of trees~\cite{geyer2005two} or even (path, tree) pairs~\cite{angelini2010tree}. Matched drawings 
require straight-line drawings of the two or more input graphs such that each
common vertex has the same $y$-coordinate in all drawings. Pairs of trees and triples of cycles always have a matched drawing~\cite{grilli2010matched}. In general, instances with no solution can be constructed from a pair of planar graphs, or even a (planar graph, tree) pair~\cite{di2009matched}. Note that matched pairs of drawings can be obtained from the MPSE embedding for every pair of graphs using the intersection line between the corresponding pairs of projection planes as the shared $y$-coordinate in the pair of matched drawings.

\smallskip\noindent{\bf Multidimensional scaling.} Consider the problem of recovering the positions (in $\mathbb{R}^p$ for some integer $p>0$, typically $p=2$) of a set of objects given their relative pairwise distances. That is, given $n$ objects with indices $1,2,\dots,n$, and given pairwise distances $\mD=[\mD_{ij}]_{i,j=1}^n$, where $\mD_{ij}$ is the observed distance between objects $i$ and $j$, we wish to compute positions $x_1,x_2,\dots,x_n\in\mathbb{R}^p$ such that $\|x_i-x_j\|\approx D_{ij}$ for all objects $i$ and $j$. The matrix $\mD$ may contain distances measured in a higher dimensional Euclidean space, or a more general metric space, may be corrupted by noise, or may just represent a measure of dissimilarity between the objects that does not come from a metric distance. (Metric) multidimensional scaling (MDS) is a well known dimensionality reduction and data visualization technique that can tackle this problem. Its goal is to find an embedding 
$\mX=[x_1,x_2,\dots,x_n]\in\mathbb{R}^p$ 
of the $n$ objects so that their pairwise distances agree with the distance or dissimilarity matrix $\mD$, by minimizing the stress function
\begin{equation}
    \label{eq:mds_stress}
    \sigma^2(\mX; \mD) := \sum_{i > j} \left( \mD_{i j} - \Vert x_i - x_j \Vert \right)^2.
\end{equation}
Minimizing the stress function~\eqref{eq:mds_stress} is typically accomplished using (stochastic) gradient descent~\cite{bottou2010large} or stress majorization~\cite{gansner2004graph}.

Note that the original formulation of multidimensional scaling is non-metric MDS. The problem was first studied in the non-metric setting by Shepard~\cite{shepard1962analysis} and  Kruskal~\cite{kruskal1964multidimensional}. Non-metric MDS recovers structure from measures of similarity, based on the assumption of a reproducible ordering between the distances, rather than relying on the exact distances. 



\smallskip\noindent{\bf Multi-view multidimensional scaling.} Given a \emph{single} pairwise distance or dissimilarity matrix $\mD$, MDS aims to find an embedding $X$ that minimizes MDS stress \eqref{eq:mds_stress}. In some applications, data is collected from multiple sources, resulting in \emph{multiple} dissimilarity matrices. The following question arises: Given $K$ dissimilarity matrices $\cD = \{ \mD^1, \mD^2,\dots, \mD^K \}$ of the same $n$ objects, how to construct a single embedding $X$ that best represents all of the dissimilarities simultaneously? The answer to this question depends heavily on how the different dissimilarity matrices are related to each other.

In the simplest of cases, the different dissimilarity matrices may correspond to different noisy measurements of the same pairwise relationship. In this case, it may be possible to estimate the true dissimilarity matrix $\mD$. For example, if noise in the observed dissimilarity matrices $\mD^k$ can be assumed to be independent and identically distributed (i.i.d.) random variables, then the average $(\mD^1 + \mD^2 + \cdots + \mD^K) / K$ is an unbiased estimate of $\mD$. It would then be possible to construct an MDS embedding using the estimate of $\mD$. Another alternative is to construct an embedding directly, by finding an embedding $X$ that best approximates all of the pairwise dissimilarities simultaneously, for example, by minimizing the functional
$$\sum_{k=1}^K \sigma^2(\mX; \mD^k)=\sum_{k=1}^K \sum_{i > j} \left( \mD_{i j}^k - \Vert x_i - x_j \Vert \right)^2.$$

Bai et. al~\cite{bai2017multidimensional} propose finding the embedding $\mX$ that minimizes the Multi-View Multidimensional Scaling (MVMD) stress function
\begin{equation*} S_{MV}(\mX,\alpha; \cD) = \sum_{k=1}^K \alpha_k^\gamma \sum_{i<j} (\mD_{ij}^k-\|x_i-x_j\|)^2, 
\end{equation*}
where the weights $\alpha = [\alpha_1,\alpha_1,\dots,\alpha_K]$ are subject to the constraints
$$\sum_{k=1}^K \alpha_k = 1,\quad 0\leq\alpha_k\leq 1,$$
and $\gamma>1$ is a fixed parameter. Here the objective is to find an embedding $X$ that minimizes a weighted sum of the MDS stress \eqref{eq:mds_stress} for the different dissimilarity matrices. The parameter $\gamma$ balances between just finding the embedding that produces the smallest single stress $\sigma^2(\mX; \mD^k)$ and assigning equal weights to all of the individual stress values. This idea, along with similar multi-view generalizations of other visualization algorithms, are implemented by Kanaan et al.~\cite{kanaan2018multiview}. 

In these approaches, the general assumption is that the different dissimilarity matrices correspond to different views of the same relationship in the data. But what if the different dissimilarity matrices truly measure different relationships in the data? In this case, an embedding of the data whose corresponding pairwise distances try to agree with all of the pairwise dissimilarities is not useful. The multi-perspective simultaneous embedding problem that we propose to solve with MPSE is different, as we ask whether it is possible to embed the data so that different perspectives $P^k(X)$ of the embedding $X$ can visualize the different dissimilarities $D^k$ simultaneously.


\smallskip\noindent{\bf 3D Reconstruction.} Our problem is also related to 3D reconstruction from a collection of 2D images. This problem has been widely studied in different settings, including reconstructing the underlying real 3D structure from large collections of 2D photos ~\cite{snavely2008modeling}. More restricted variants are even closer to our setting~\cite{koutsoudis2014multi, queau2017dense}. Note however, that in our problem we have a constant number of inputs (distance matrices or graphs) and the projections we anticipate can be fixed or computed as a part of the optimization. 
Our proposed problem and its solution can be viewed also as a step of the  structure-from-motion (SfM)  problem~\cite{ozyesil2017survey}. The SfM problem is one of the central problems in computer vision which aims to recover the 3D structure of a scene from a set of its projected 2D images. We relate our problem to the SfM problem by assuming some fixed points in the 3D scene which can be estimated on the corresponding 2D projections via some feature extraction algorithm such as SIFT~\cite{lowe1999object}. Once these points are estimated, we can measure the pairwise distance matrices for these points and apply the MPSE algorithm to find the locations of these points in 3D.

\subsection{Our Contribution}
The main contribution in this paper is the introduction of the multi-perspective simultaneous embedding problem and a generalization of MDS to multiple distance matrices to solve it. This is at the core of the proposed MPSE method for visualizing the same dataset/graph in 3D from several different views, each of which captures a different set of distances/relationships. We consider two main variants: one in which each of the different distances/relationships is associated with a specific 2D projection plane, and the other where computing the projection planes is also part of the optimization. Extensive quantitative experiments show that the method scales well with increasing number of data points and increasing number of projections.



\section{Multi-Perspective, Simultaneous Embedding}
\label{sec:fixed}

Our proposed MPSE algorithm is a generalization of the standard (metric) MDS problem. 
The setting of MDS and its corresponding optimization function, which is called stress function is discussed in Section~\ref{sec:previous_work}. We remark that  if the minimum of the MDS stress function is zero, then the objects can be positioned in $\mathbb{R}^p$ so that their pairwise distances exactly represent the pairwise dissimilarities of  $\mD$. If the minimum of the MDS stress function is nonzero but not too large, the minimizer of the MDS stress function provides an approximate way to visualize the dissimilarities in $\mathbb{R}^p$.

Suppose now that instead of a single pairwise dissimilarity matrix $\mD$, we observe multiple pairwise dissimilarities matrices $\cD = \{ \mD^1, \mD^2,\dots, \mD^K \}$ for the same set of $n$ objects. It is natural to ask if MDS can be generalized to produce an embedding so that the dissimilarities $\mD^1, \mD^2,\dots, \mD^K$ can be visualized by the relative positions of the objects.
If it is assumed that the dissimilarities in $\cD$ are no more than approximations of some hidden true dissimilarity matrices $\mD_{true}^1, \mD_{true}^2, \dots, \mD_{true}^K$, then simple generalizations of MDS exist, as described in section \ref{sec:previous_work}. In all of these generalizations, an embedding $\mX$ is produced so that the distances $\|x_i-x_j\|$ are as close as possible to all of the dissimilarity measurements $\mD^1_{ij}, \mD^2_{ij},\dots, \mD^K_{ij}$. Nonetheless, this is an undesirable assumption if the goal is to visualize multiple \emph{distinct} relationships between the same objects. In particular, the dissimilarity measurements $\mD^1_{ij}, \mD^2_{ij},\dots,\mD^K_{ij}$ can be so different that no embedding $X$ can produce pairwise distances $\|x_i-x_j\|$ that help visualize all of these relations in a meaningful way.
 

Our generalization of MDS is inspired by the problem of 3D reconstruction from multiple 2D images. 
We first explain our motivation with an example based on a sculpture of James Hopkins, which illustrates how different the same object can look when observed from a different viewpoints. 
Depending on the direction from which the viewer sees the sculpture, the viewer will see a different digit '1' or '2' or '3'. 
For our experiments, we recreated the ``1, 2, 3" sculpture and  uniformly at random fixed a set of points in the three dimensions that produces the same visual effect when plotted. 
The set of points forms a figure '1' when viewed from the $x$-axis, it forms a figure '2' when viewed 60 degrees towards the $y$-axis, and finally as a figure '3' when viewed another 120 degrees towards the $y$-axis.
Suppose now that the 3D structure is unknown, and the goal is to construct it using only the information from these 3 images.
That is, we know the 3 distance matrices $\mD^1, \mD^2, \mD^3$ measuring the distances between the same set of keypoints in each of these images, and the goal is to find an embedding $\mX$ so that
\begin{equation*} 
\mD^k_{ij} \approx \|P^k(x_i)-P^k(x_j)\|, \quad k=1,2,3,\quad i,j=1,2,\dots,n,
\end{equation*}
where $P^1,P^2,P^3$ are the projections that map keypoints in 3D to their corresponding images. For this example one can take the following projection matrices:
\begin{equation}
\label{eq:60deg_proj}
\mP^1 =
\begin{bmatrix}
     0 & 1 & 0 \\
     0 & 0 & 1
\end{bmatrix},
\mP^2 =
\begin{bmatrix}
     \frac{\sqrt{3}}{2} & \frac{1}{2} &  0 \\
    0 & 0 & 1 \\
\end{bmatrix},
\mP^3 =
\begin{bmatrix}
    \frac{\sqrt{3}}{2} & -\frac{1}{2} & 0\\
    0 & 0 & 1 \\
\end{bmatrix}.
\end{equation}
We can ask two questions: 1) given the dissimilarity matrices and the projections, can we recover the embedding $\mX$? 2) given the dissimilarity matrices, can we recover the projection matrices and the embedding $\mX$? Clearly the latter question is harder to answer.  In both cases, we suggest a natural generalization of the MDS stress function
\begin{equation*} 
\sum_{k = 1}^3 \sum_{i > j} \left( D^k_{i j} - \Vert P^k(x_i) - P^k(x_j) \Vert \right)^2 .
\end{equation*}
This can be naturally generalized to any number of distance (dissimilarity) matrices and projections. Even if the multiple dissimilarity measures at hand are not related to each other in such a geometric way, a similar idea may still provide a useful visualization of the multiple relations involved. 

Motivated by these examples, we generalize the MDS stress function in the following way: Given $n$ objects with $L$ distinct dissimilarity matrices ${D^l}$, the multi-perspective MDS stress function is defined as
\begin{equation}
\label{eq:multi-perspective-mds-stress}
\begin{split}
    S(\mX,\cP;\cD) & = \sum_{k=1}^L \sigma^2(P^{(k)}(\mX);\mD^{(k)})
    \\ & = \sum_{k=1}^L \sum_{i > j} \left( \mD^{(k)}_{i j} - \Vert P^{(k)}(x_i) - P^{(k)}(x_j) \Vert \right)^2.
\end{split}
\end{equation}
Here, $\cP =[P^{(1)},P^{(2)},\dots,P^{(K)}]$, where ${P^k:\mathbb{R}^p\to\mathbb{R}^q}$ are the mappings (projections). We write $P^{(k)}(X) = [P^{(k)}(x_1),P^{(k)}(x_2,\dots,P^{(k)}(x_n)]$. We mainly focus on the special case of linear orthogonal mappings, which correspond to orthogonal projections. There are two different ways in which the stress function \eqref{eq:multi-perspective-mds-stress} can be minimized: 1) Given a fixed set of mappings ${P^l}$, we find a minimizer $\mX\in\mathbb{R}^{n\times p}$; 2) we find a pair $(\mX, \cP)$ minimizing the stress, where $\mX\in\mathbb{R}^{p\times n}$ and the mappings $P^k$ for $1 \le k \le L$ are restricted to a class of functions (such as linear orthogonal transformations from $\mathbb{R}^p$ to $\mathbb{R}^q$).

 
 

\subsection{MPSE with Fixed Perspectives}
\label{sec:mpse_fixed}
We first consider minimization of the multi-perspective MDS stress function \eqref{eq:multi-perspective-mds-stress} with respect to the embedding $\mX$ only:
\begin{equation}
    \label{eq:problem_fixed}
    \underset{x_i\in\mathbb{R}^p}{\mathrm{minimize}} \;
    S(\mX,\cP; \cD)
\end{equation}
In this setup, the perspective mappings $P^{(1)},\dots,P^{(K)}:\mathbb{R}^p\to\mathbb{R}^q$ are predetermined and fixed. We assume that these mappings are differentiable, so that a solution to \eqref{eq:problem_fixed} may be attainable using a gradient descent scheme. 

We remark that the objective function for MPSE with fixed projections in \eqref{eq:problem_fixed} is differentiable. Since the multi-perspectives MDS stress function \eqref{eq:multi-perspective-mds-stress} is generally non-convex, minimization to a global minimum is not guaranteed by a vanilla implementation of gradient descent. In addition, computing the full gradient of \eqref{eq:multi-perspective-mds-stress} at each iteration is expensive for large data sets and might make the algorithm infeasible. Thus coming up with a fast and accurate solution is one of the main contributions in this paper. 
In section \ref{sec:algorithms}, we describe a stochastic gradient descent scheme with adaptive learning rate for efficient solution to this problem.

\subsection{MPSE with Varying Projections}
\label{sec:mpse_varying}

We again consider minimization of the multi-perspective MDS stress function \eqref{eq:multi-perspective-mds-stress}, but we no longer assume predetermined and fixed perspective mappings $\cP$. Our goal is then to find both the embedding $\mX$ and the perspective mappings $\cP$ that best capture the given distance matrices $\cD$. In order to do this, a suitable parametric family $\cO$ of $C^1(\mathbb{R}^p,\mathbb{R}^q)$ perspective mappings must be defined. The optimization problem becomes
\begin{equation}
    \label{eq:problem_varying}
    \underset{ \mX \in\mathbb{R}^{p\times n},\;P^{(k)}\in\cO} {\mathrm{minimize}} \; S(\mX, \cP; \cD)
\end{equation} 

For ease of exposition, we only consider the set of orthogonal projections for the family of perspective mappings. 
Section~\ref{sec:relevant_math}.
By $\mathbb{O}^{3\times 2}$ we denote the family of orthogonal projections from $\mathbb{R}^3$ to $\mathbb{R}^2$.  Then, each perspective mapping can be parameterized by an orthogonal $p \times q$ matrix $Q$. The perspective mapping $P^{(Q)}$ is given by $x \mapsto Qx$. Note that the perspective mappings vary smoothly with respect to changes in the parameters, in the sense that the map $Q \mapsto P^{(Q)}(x)$ is $C^1$ for every $x\in\mathbb{R}^p$. The optimization problem \eqref{eq:problem_varying} can then be rewritten as
\begin{equation}
   \label{eq:problem_orthogonal}
    \underset{\mX\in\mathbb{R}^{p \times n},\;Q^{(k)}\in\mathbb{O}^{3\times 2}} {\mathrm{minimize}} \; \sum_{k=1}^n \sigma^2(Q^{(k)} \left( \mX\right); \cD)
\end{equation}


We remark that the objective function of MPSE with varying projections in \eqref{eq:problem_orthogonal} is also differentiable. However, since the set $\mathbb{O}^{3\times2}$ of $3$ by $2$ orthogonal matrices is not a subspace of $\mathbb{R}^{3\times2}$, minimizing \eqref{eq:problem_orthogonal} with respect to $Q_l$ cannot be accomplished via gradient descent. Instead, we make use of projected gradient descent, where $Q_l$ is updated by first moving towards the direction of steepest descent, and then projecting back onto the set $\mathbb{O}^{3\times 2}$. If $A\in\mathbb{R}^{3\times 2}$ matrix, then the projection of $A$ onto $\mathbb{O}^{3\times 2}$ is the matrix $\Pi(A) \in \mathbb{O}^{3\times 2}$ that minimizes $\| A - Q \|_F$ among all $Q \in \mathbb{O}^{3\times 2}$. There is a simple way to compute $\Pi(A)$: if $U \Sigma V^T$ is the reduced singular value decomposition of $A$, then $\Pi(A) = U V^T$. 

\subsection{Some Extensions}
\label{sec:extensions}
The MDS stress function can be extended to general graph structures. That is, we do not require that a pairwise dissimilarity relation exists between every pair of nodes. Moreover, we assume we are given a weighted pairwise relation, where some edges contribute to the stress function more than the others. To make stress values more meaningful, it is common to work with a normalized MDS stress function. The normalized MDS stress function is
\begin{equation}
    \label{eq:normalized-mds-stress}
    \tilde{\sigma}^2(\mX, \mD) = \frac{\sum_{(i,j) \in E} w_{ij} \left( \mD_{i j} - \Vert (x_i) - (x_j) \Vert \right)^2}{\sum_{(i,j) \in E} w_{ij} (\mD_{i j}) ^2} .
\end{equation}
where $E$ is the set of edges in the graph for which a dissimilarity relation exits and $w_{ij}$ is the weight corresponding to edge $(i,j)$.

We also generalize MPSE to weighted graphs, where the normalized MPSE stress function becomes
\begin{equation}
\label{eq:normalized-mpse-stress}
    \tilde{S}^2(\mX,\cP; \cD) =  \frac{1}{K} \sum_{k=1}^K \tilde{\sigma}^2(P^{(k)}(\mX), \mD^{(k)}),
\end{equation}
where $\mP^{(k)}(X) = [\mP^{(k)}(x_1), \mP^{(k)}(x_2),\dots, \mP^{(k)}(x_N)]$ for each $k=1, 2, \dots, K$.

\section{Algorithms}
\label{sec:algorithms}

In this section we discuss our proposed algorithms for solving the optimization problems for MPSE with fixed projections \eqref{eq:problem_fixed} and MPSE with varying projections \eqref{eq:problem_varying}. 
In our experiments, we found that convergence to a global minimum of \eqref{eq:multi-perspective-mds-stress} is unlikely using basic gradient descent algorithms. 
We found that a combination of smart initialization and stochastic gradient descent with adaptive learning rate usually produces better results than the vanilla version of the gradient descent. The benefits of using stochastic gradient are twofold: first, it is faster than the vanilla version of the gradient descent, and second, as observed by others~\cite{zheng2018graph}, it helps avoid local minima.

Given a dissimilarity matrix $\mD$, let $\xi \sim \Xi(\mD, c)$ be a random variable returning a dissimilarity matrix $\xi$, where for each $(i, j)$, $1 \le i, j \le n$ we include $\mD_{ij}$ with probability $c$ and otherwise set  $\xi_{ij} = 0$.
Let $\sigma^2(\mX,\xi)$ and $\nabla_{\mX} \sigma^2(\mX,\xi)$ be unbiased estimates of the MDS stress and its gradient at $\mX$. Then, the multi-perspective MDS stress \eqref{eq:multi-perspective-mds-stress} is approximated by
\begin{equation*}
S_\mathbb{\xi}^2(\mX,\cP; \cD) = S^2(\mX, \cP, \mathbb{\xi})
\end{equation*}
where $\mathcal{\xi}=[\xi^1,\dots,\xi^K]$ and $\xi^{(k)} \sim \Xi ({\mD^{(k)},c})$. The gradients $\nabla_{\mX} S_\mathbb{\xi}^2(\mX,\cP;\cD)$ and $\nabla_{Q^{(k)}} S_\mathbb{\xi}^2(\mX,\cP;\cD)$ are similarly computed. Note that all these can be computed simultaneously with one pass through the edge list.

We use an adaptive gradient descent framework from \cite{malitsky2019adaptive} with the following adaptive scheme for the learning rate:
\begin{equation}
\label{eq:learning_rate_update}
\mu_{\mX} ({\mX}_t,{\mX}_{t-1},\xi_{t}) = \frac{(\mX_t - \mX_{t-1})^T(\nabla_{\mX} S_{\mathbb{\xi}_t}^2 (\mX_t - \nabla_{\mX} S_{\mathbb{\xi}_T}^2(\mX_{t-1}))}
{ \left \Vert \nabla_{\mX} S_{\mathbb{\xi}_t}^2(\mX_t) - \nabla_{\mX} S_{\mathbb{\xi}_t}^2(\mX_{t-1}) \right \Vert^2} 
\end{equation}

The adaptive learning rate for the projection matrices $\mu_Q(t)$ is computed in a similar fashion.

We summarize the resulting algorithm for  MPSE with fixed projections\eqref{eq:problem_fixed} in Algorithm~\ref{algo:fixed}. We remark that the problem is non-convex and there is no guarantee for convergence. However, our extensive experiments show that with smart initialization, or with multiple runs of the algorithm (with different random initialization), the algorithm consistently finds good solutions.
The idea behind the smart initialization is that even though the problem is non-convex, around the global optimum the problem is well-behaved and so starting close to the global optimum we can apply gradient descent.
\begin{algorithm}
\caption{MPSE with fixed projections (problem \eqref{eq:problem_fixed})}
\label{algo:fixed}
\begin{algorithmic}
\Require Dissimilarity matrices: $\cD = \{ \mD^1, \dots, \mD^K\}$, initial embedding $\mX_0$, initial learning rate $\mu_0$, stochastic constant $c$, iteration number $T$.\\
\For{$t=1,2,\dots,T$}
\State $\xi_t \sim \Xi(\mD, c)$
\State $\mX_t = \mX_{t-1} - \mu_{t-1} \nabla_{\mX} S^2_{\xi_t} (\mX_t,\mathbf{P};\mathbf{D})$ 
\State $\mu_t = \mu(\mX_{t-1}, \mX_t,\xi_t)$
\EndFor
\Ensure $\mX_T$
\end{algorithmic}
\end{algorithm}

Next, we describe the algorithm for MPSE with variable projections \eqref{eq:problem_varying} and summarize it in Algorithm~\ref{algo:varying}. We remark that this problem is more complex then MPSE with fixed projections. Variable projections offer more degrees of freedom, at the expense however of non-convex optimization  and non-trivial local optima. Extensive experiments in this setting also show that the algorithm works well in practice. 
\begin{algorithm}
\caption{MPSE with varying perspectives (problem \eqref{eq:problem_varying})}
\label{algo:varying}
\begin{algorithmic}
\Require Dissimilarity  matrices: $\cD = \{ \mD^1, \dots, \mD^K\}$, initial embedding and perspective parameters $\mX_0$ and $\mathbf{Q}_0$, initial learning rates $\mu_{\mX}$, and $\mu_Q$, stochastic constant $c$, iteration number $T$.
\For{$t=1,2,\dots,T$}
\State $\xi_t \sim \Xi(\mD, c)$
\State Compute $\nabla S^2_\xi (X_{t-1},\bold{Q}_{t-1})$ and $\nabla S^2_\xi (X_{t},\bold{Q}_{t})$
\State Compute $\mu_t$.
\State $\mX_t = \mX_{t-1} - \mu_{X,0} \nabla_{\mX} S^2_\xi (\mX,\bold{Q}_t)$.
\State $\bold{Q}_t = \Pi(\bold{Q}_t - \mu_{\bold{Q}_t} \nabla_Q S^2_\xi)$
\State $\mu^X_t = \mu(\mX_{t-1}, \mX_t,\xi_t)$ 
\State $\mu^Q_t = \mu(\mathbf{Q}_{t-1},\mathbf{Q}_t,\xi_t)$ 
\EndFor
\Ensure $X_\textrm{final}$, $\bold{Q}_\textrm{final}$.
\end{algorithmic}
\end{algorithm}

Algorithm~\ref{algo:initialization} summarizes our proposed smart initialization of $\mX_0$ and $\mQ_0$. We have found that a preliminary computation of the optimal 'combined' MDS embedding of the dissimilarity relations works best. Finally, we remark that the standard practice of multiple runs with different random initialization also helps avoid local minima.
\begin{algorithm}
\caption{Initialization for Algorithm \eqref{algo:varying}}
\label{algo:initialization}
\begin{algorithmic}
    \Require Dissimilarity set $\mathcal{D}$, random initial embedding and perspective parameters $X_0$ and $\mathcal{Q}_0$, initial learning rates $\mu_X$, and $\mu_Q$, stochastic constant $c$, iteration number $T$.\
    \State $D_{ij}^2= \frac{d_1}{d_2 K}\sum_{k=1}^K (D_{ij}^{(k)})^2$
        \Comment{Combine dissimilarities.}
    \For{$t=1,2,\dots,T$}
        \State $\xi_t \sim \Xi(D,c)$
        \State $\mX_t = \mX_{t-1} - \mu_{t-1} \nabla_{\mX} S^2_{\xi_t} (\mX_t,\mP;\mD)$ 
        \State $\mu_t = \mu(\mX_{t-1}, \mX_t,\xi_t)$
    \EndFor
        \Comment{Compute MDS embedding of $\mD$ on $\mathbb{R}^{d_1}$:}
    \State $\mX = \mX_t$.
    \For{$t=1,2,\dots,T$}
        \State $\xi_t \sim \Xi(\mD,c)$
        \State $\mQ_t = \mQ_{t-1} - \mu_{t-1} \nabla_{\mQ} S^2_{\xi_t} (\mX_t,\mQ_t;\cD)$ 
        \State $\mu_t = \mu(\mX_{t-1}, \mX_t,\xi_t)$
    \EndFor
        \Comment{Compute optimal perspective parameters given embedding $\mX$:}
    \State $\mX = \mX_t$.
    \Ensure $\mX_\textrm{final}$, $\mQ_\textrm{final}$.
\end{algorithmic}
\end{algorithm}

\section{Experimental Evaluation}
\label{sec:numerical_experiments}

\begin{figure*}
\includegraphics[width=0.195\linewidth]{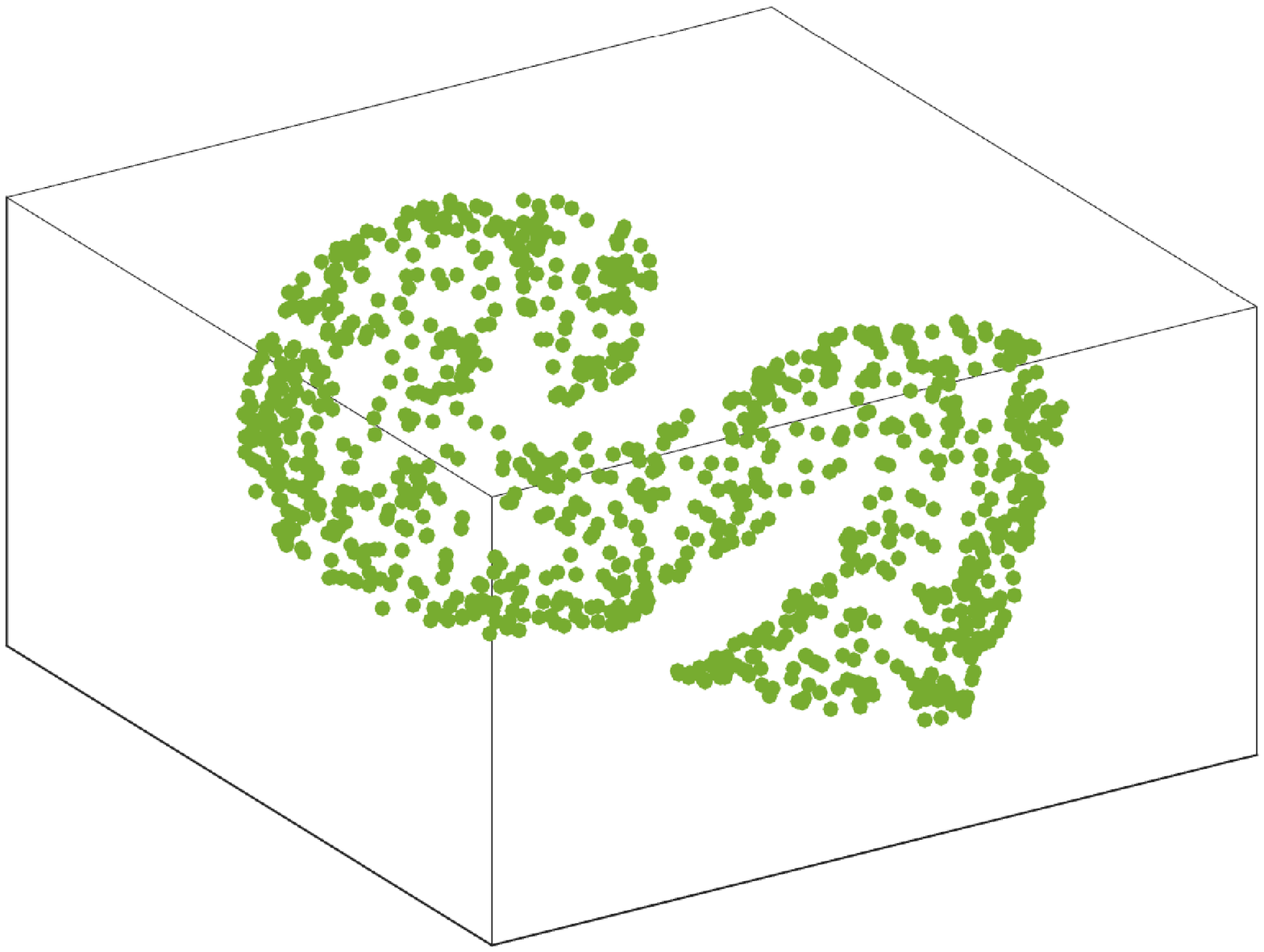}
\includegraphics[width=0.195\linewidth]{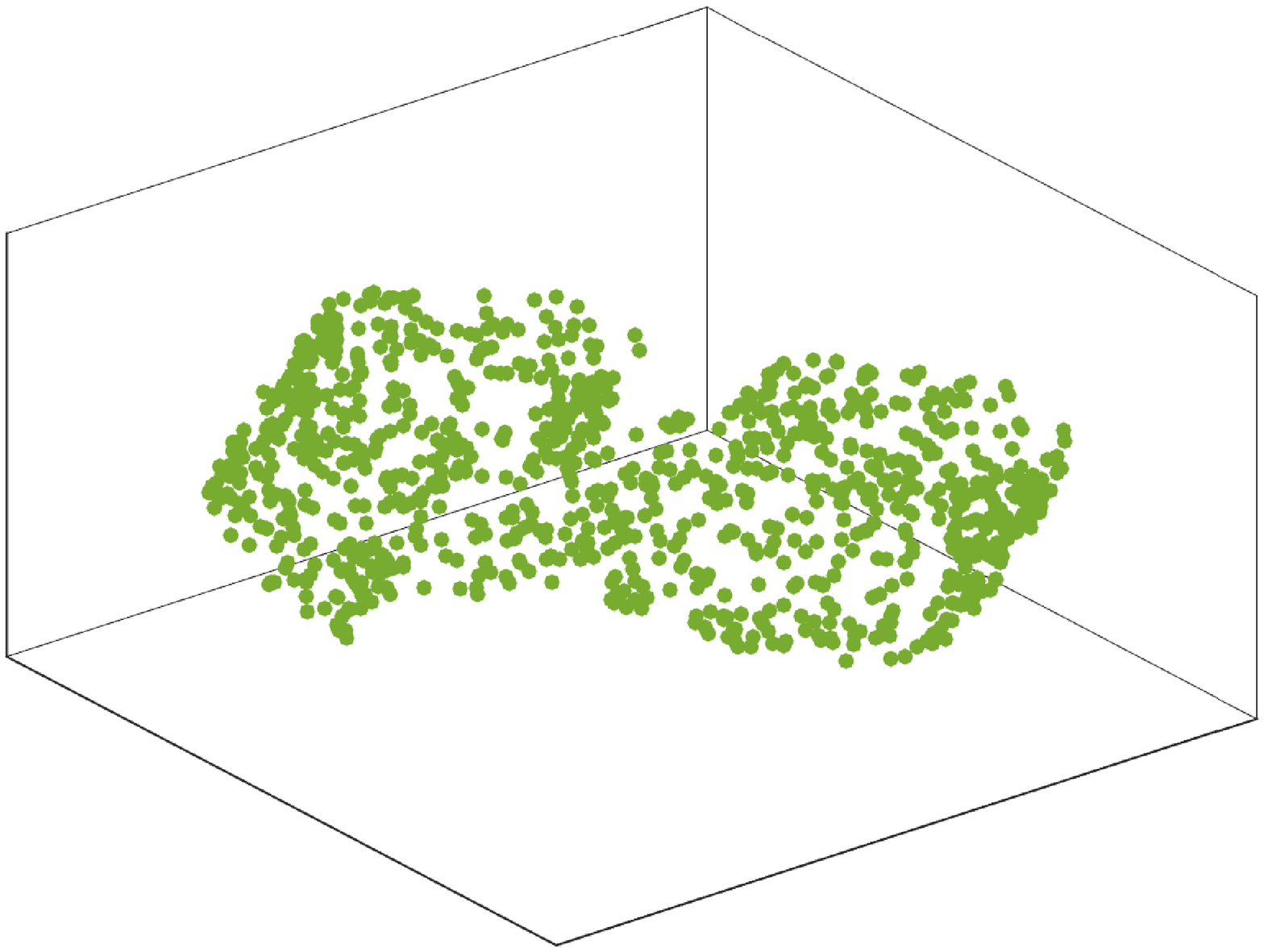}
\includegraphics[width=0.195\linewidth]{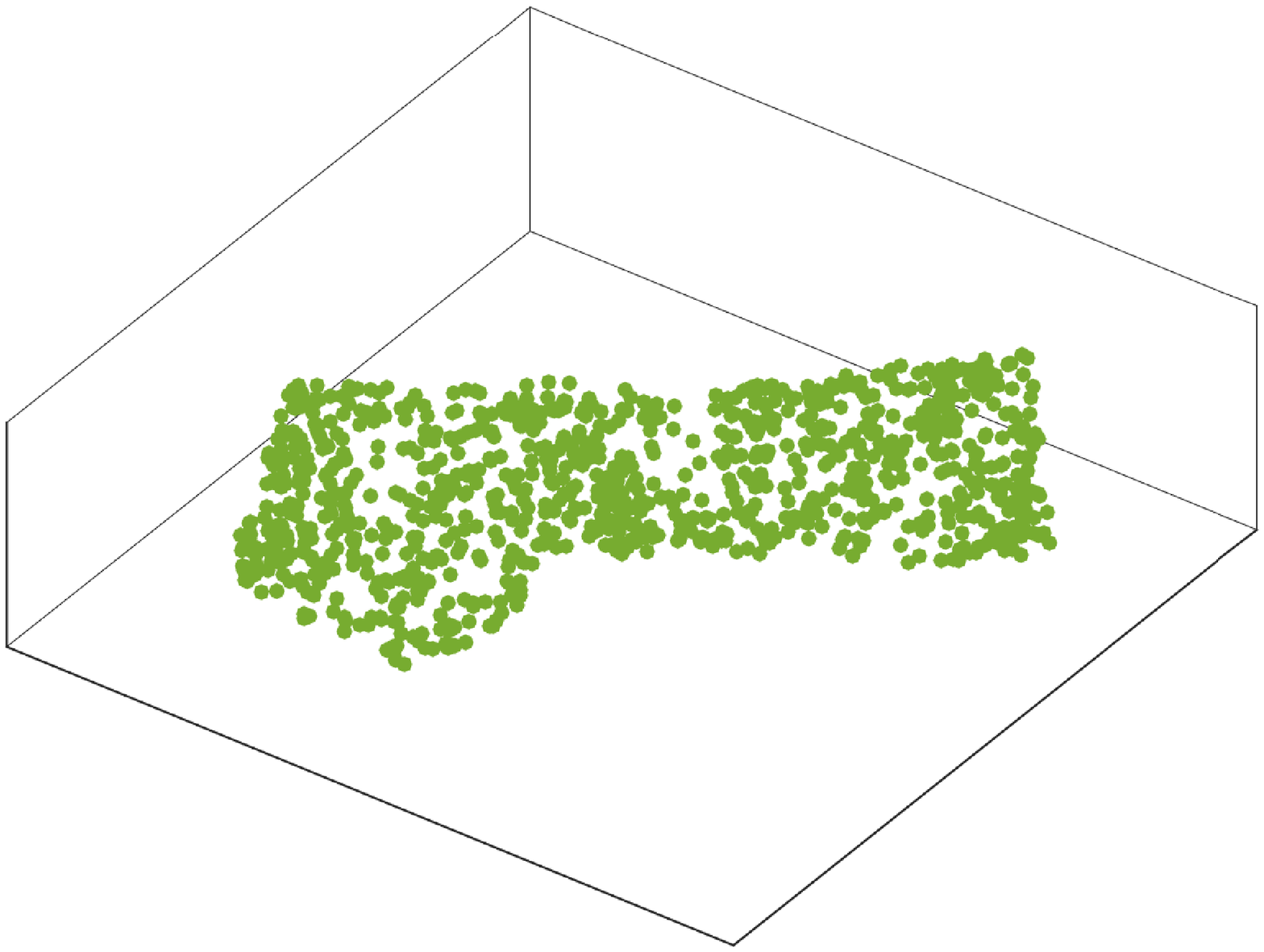}
\includegraphics[width=0.195\linewidth]{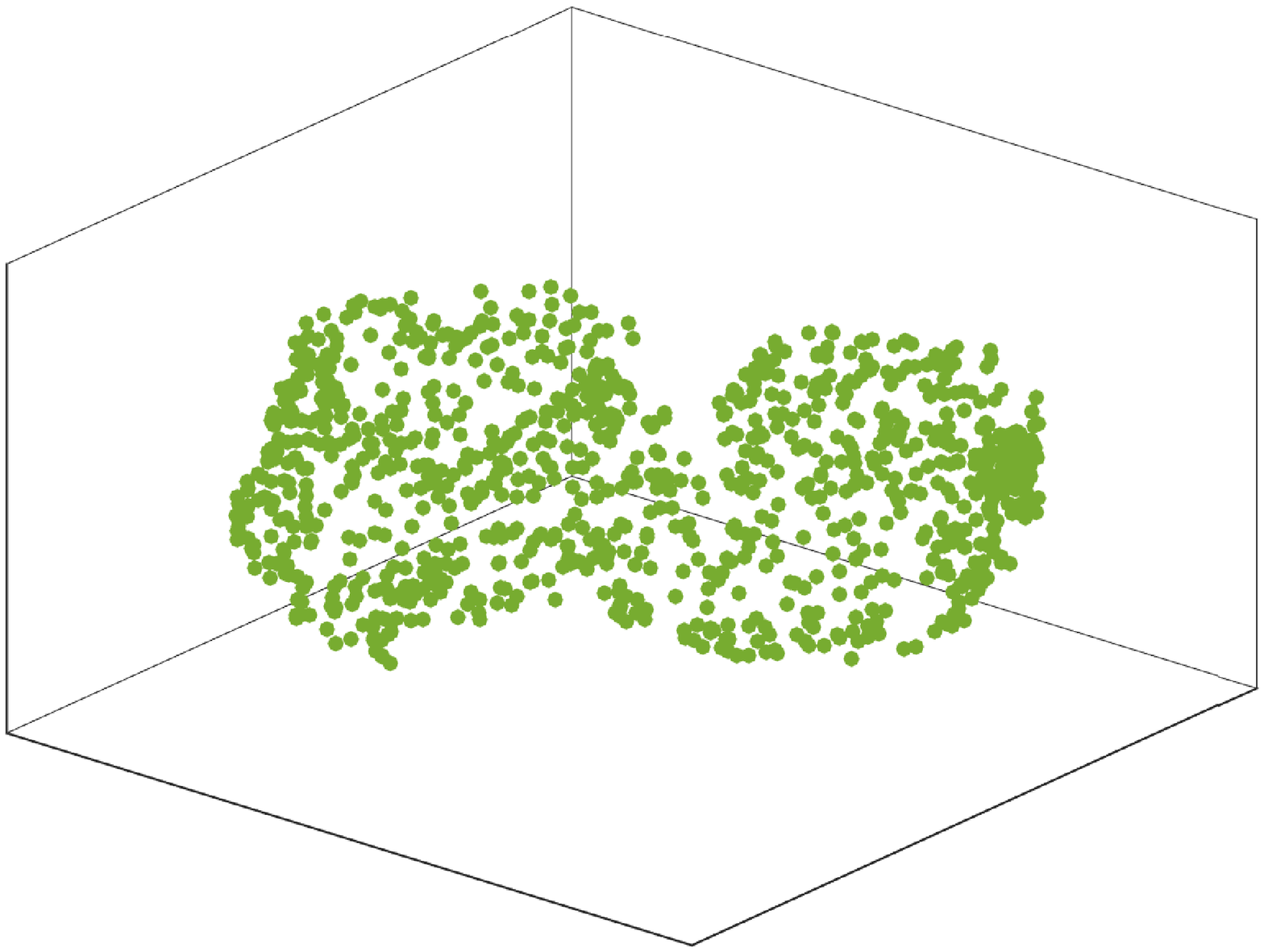}
\includegraphics[width=0.195\linewidth]{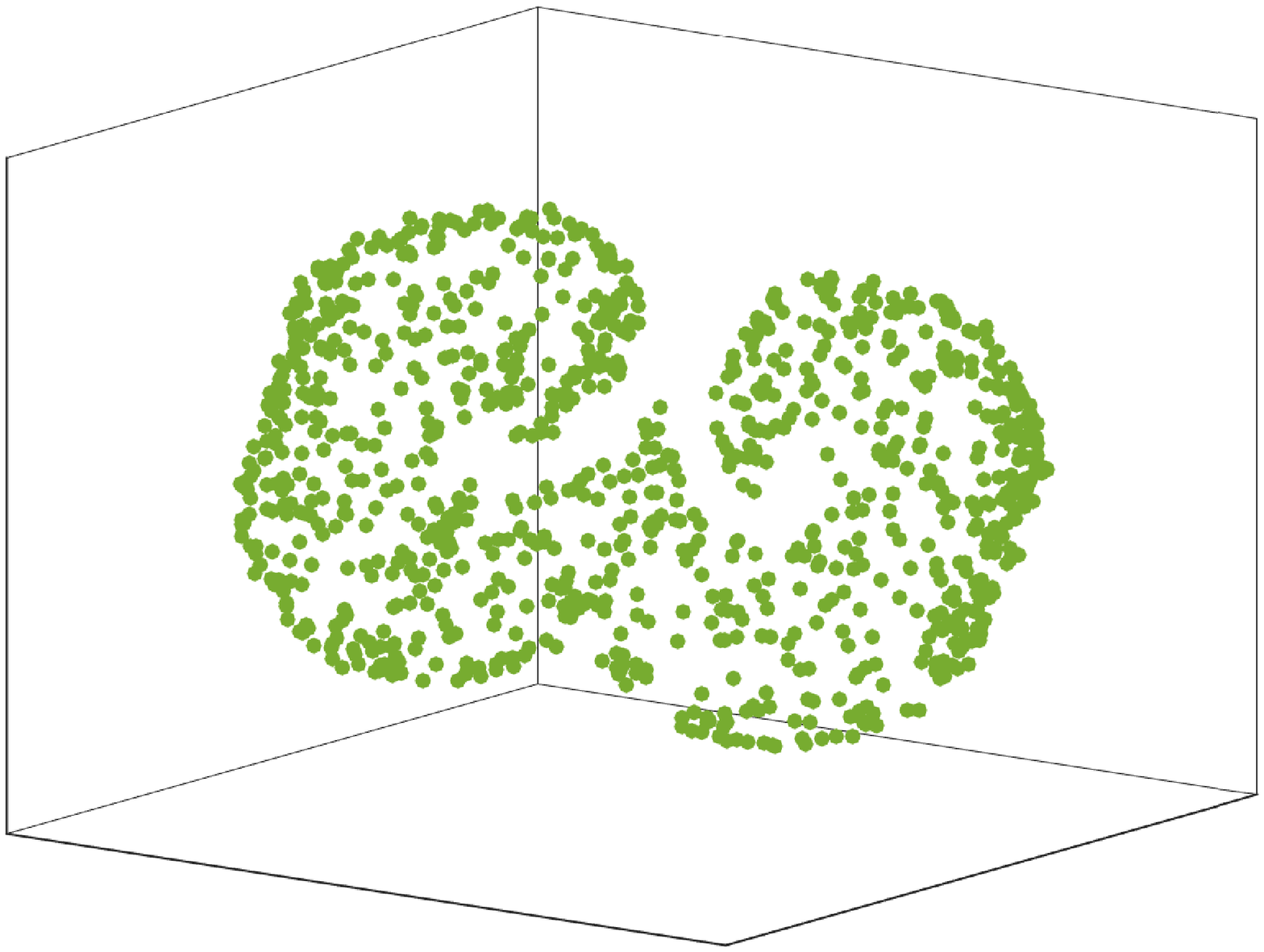}
\caption{The output of Algorithm~\ref{algo:varying} on the distance matrices of the One-Two-Three dataset form Fig.~\ref{fig:123_one_orig}. The first subfigure shows the projection which recovers digit 2, the third one corresponds to digit 1, and the last one corresponds to digit 3. The second and fourth shows  the  rotational transition in 3D between 2 and 1 and between 1 and 3. The final MPSE stress for the One-Two-Three dataset with varying projections (Algorithm \ref{algo:varying}) is 0.07, with projection-wise stress values of 0.04, 0.08 and 0.08.}
\label{fig:123_varying_results}
\end{figure*}

In this section we experimentally evaluate the proposed MPSE algorithms: with fixed projections and with variable projections. Since the problem that we study is new,  there are no algorithms to compare against; instead, we create benchmark datasets that we believe can demonstrate and test the MPSE algorithms. For each dataset, we first describe how it was created and show the outputs of MPSE. Note that Algorithm~\ref{algo:varying}, similar to the regular MDS, computes results invariant to global rotation/reflection and there is no way to recover the exact orientation of the embedding without additional information.

\subsection{One-Two-Three Dataset}
\label{sec:one_two_three}

\begin{figure}[H]
\includegraphics[width=0.32\linewidth]{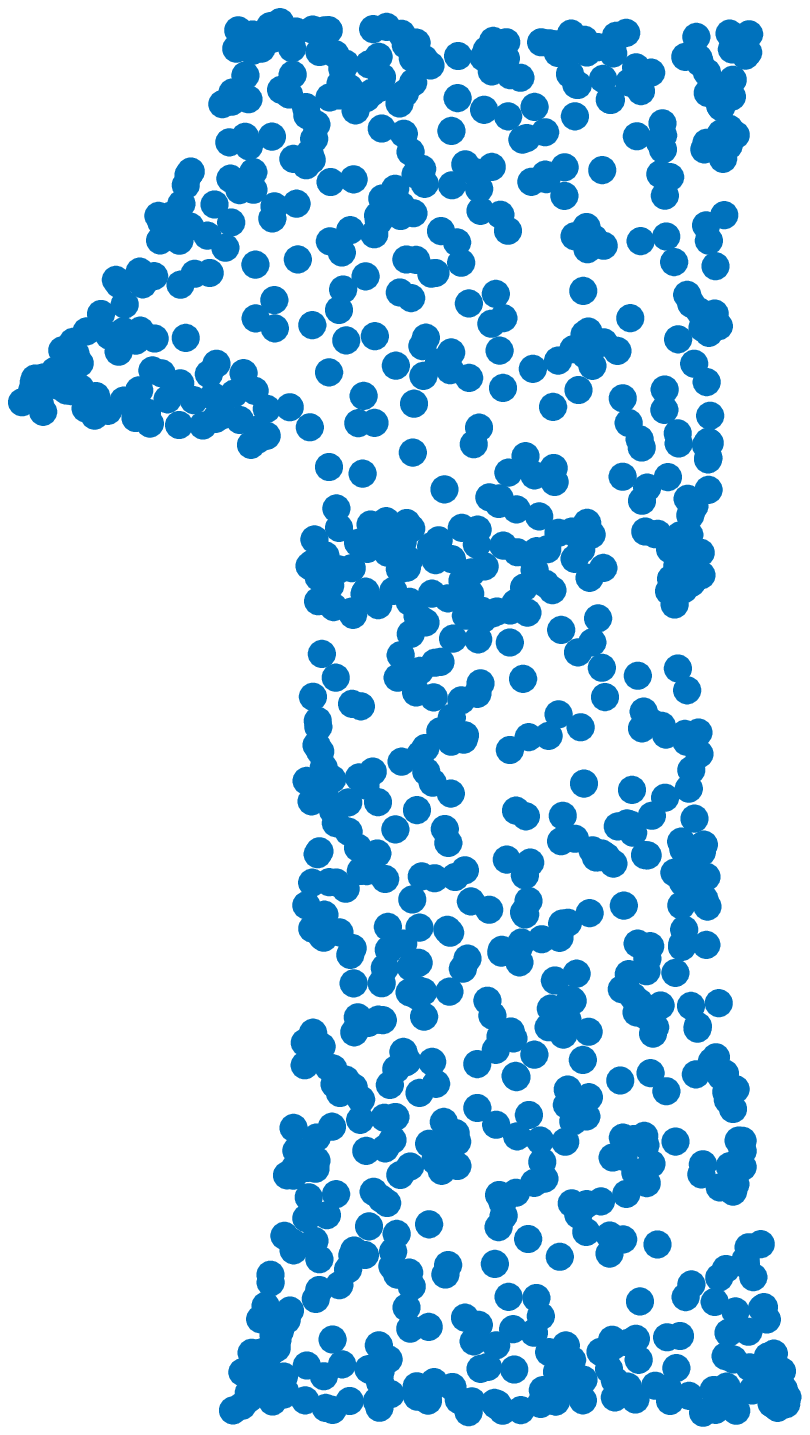}
\includegraphics[width=0.32\linewidth]{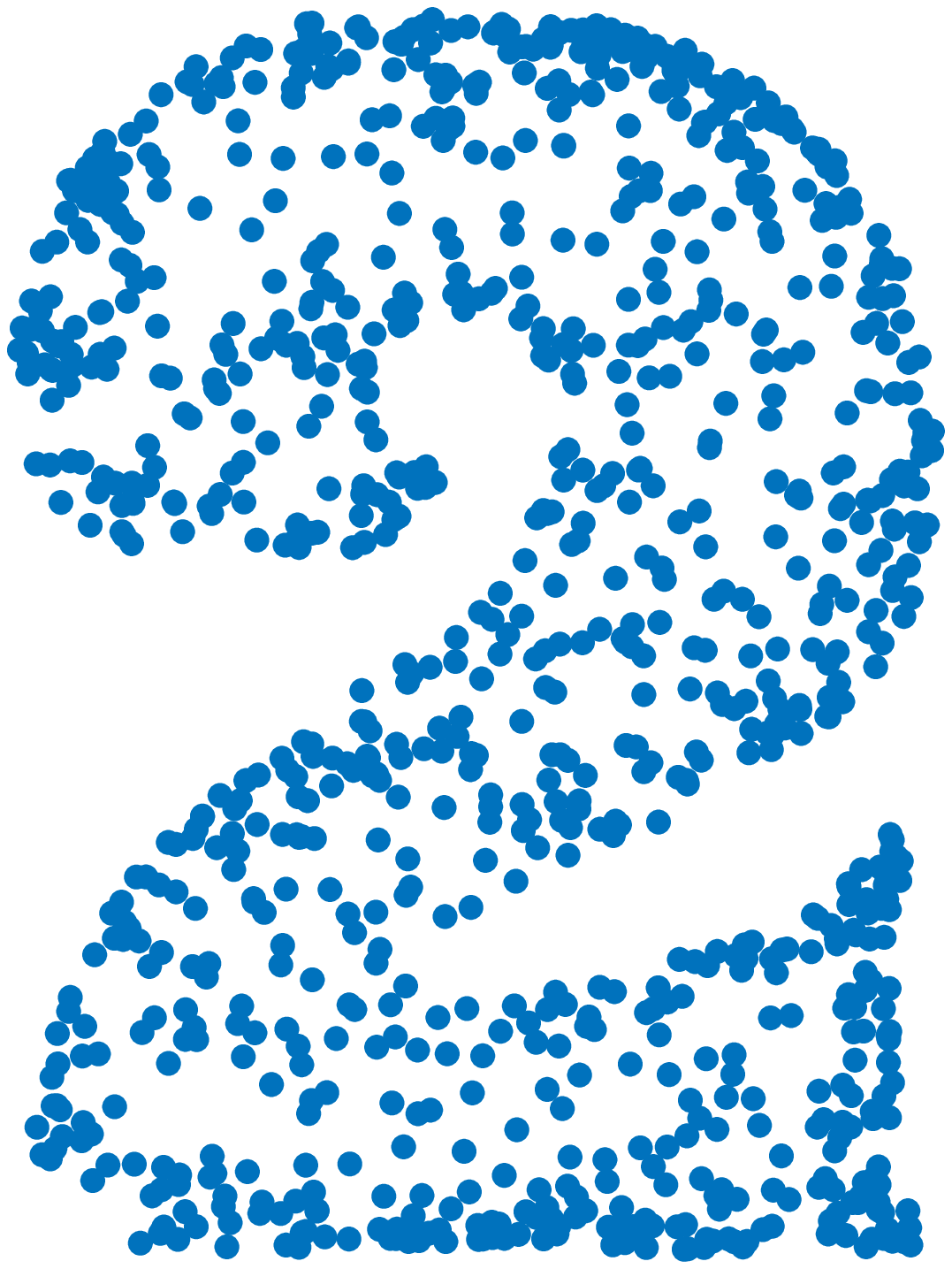}
\includegraphics[width=0.32\linewidth]{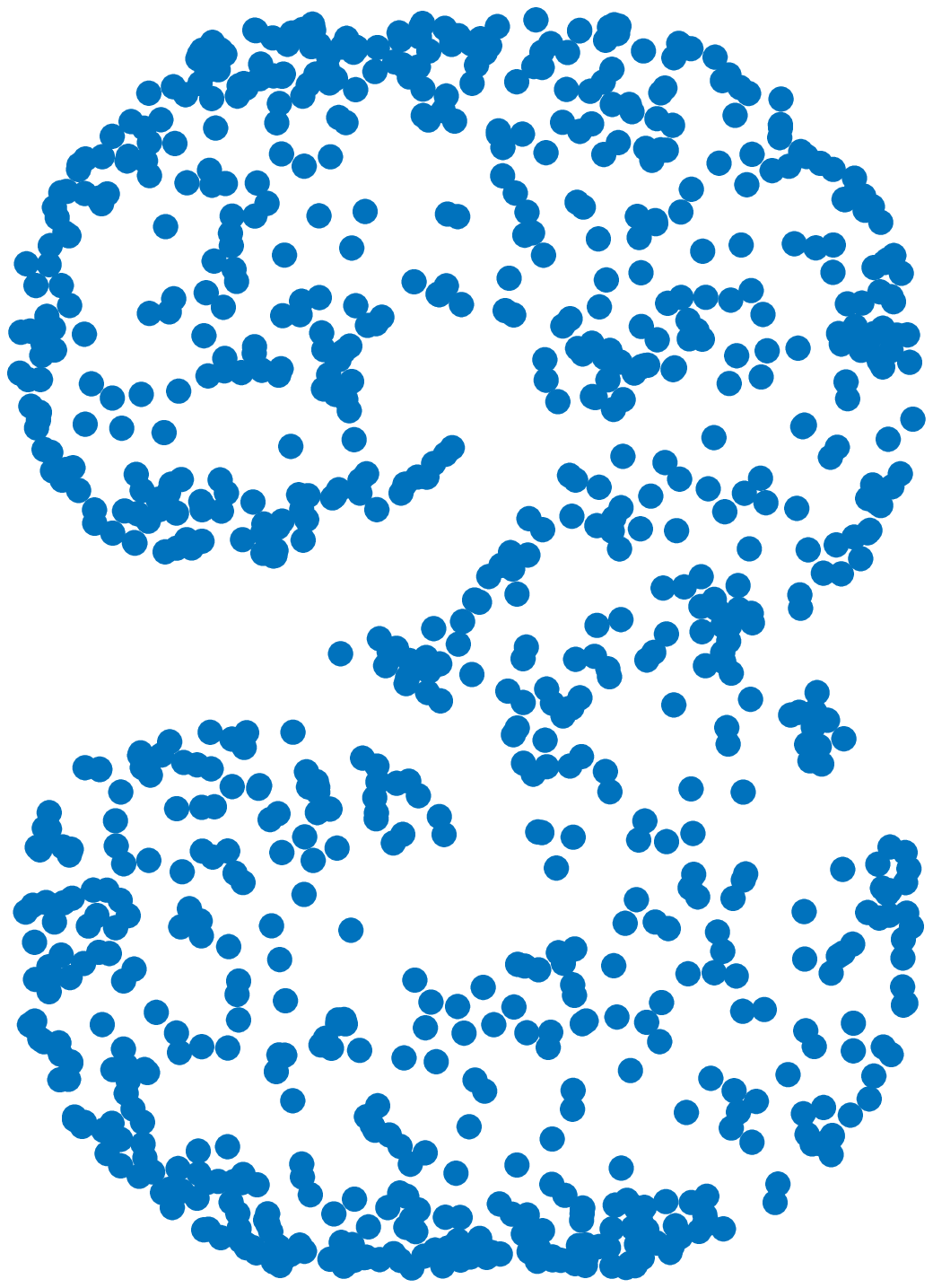}
\caption{One-Two-Three Dataset: each subfigure contains 1000 points sampled from the original images of digits 1, 2 and 3. The points are labeled from top to bottom to preserve their correct positions.}
\label{fig:123_one_orig}
\end{figure}

In the introduction we motivated the MPSE problem using the sculpture ``1, 2, 3" by James Hopkins. To generate the dataset for Fig.~\ref{fig:teaser} we attempted to reverse-engineer the sculpture as follows: We created 2D visualizations of the 3 projections by creating images of the digits 1, 2 and 3 and  sampled $1000$ points from each; see Fig.~\ref{fig:123_one_orig}. 

Then we computed the 2D distances for each set of points and fed those as  input to the MPSE algorithm, aiming to recover the 3D figure of the sculpture. The goal is to see whether it is possible to place $1000$ points in 3D so that it would look like digits 1, 2 and 3 when seen from different viewpoints. 

We first run Algorithm~\ref{algo:fixed} for the distance matrices created from the One-Two-Three dataset; see Fig.~\ref{fig:123_one_orig}. For the algorithm we used the following parameters: maximum number of iterations $T = 100$,  fixed projections $\mP^1, \mP^2$ and $\mP^3$ from \eqref{eq:60deg_proj}, starting learning rate $\mu_0 = 1$ and stochastic constant $c = 0.01$ with random initialization. The results are shown in Fig.~\ref{fig:teaser} and indicate that the algorithm successfully placed the points in 3D such that one can see digits 1, 2 and 3 (see the first, third and fifth subfigures of Fig.~\ref{fig:teaser}). The quality of the MPSE embedding in Fig.~\ref{fig:teaser} is also quantified with low stress values:
overall stress 0.001, with projection-wise  stress values of 0.0009, 0.0008 and 0.0011.

Next, we run Algorithm~\ref{algo:varying} for the same One-Two-Three dataset, with the following  parameters: maximum number of iterations $T = 100$, the starting learning rate $\mu_0 = 1$ and stochastic constant $c = 0.01$ with smart initialization described in Algorithm~\ref{algo:initialization}. The results are shown in Fig.~\ref{fig:123_varying_results}. We remark that for the MPSE with varying projections the results are correct, up to a global rotation and reflection. The algorithm again successfully placed the points in 3d such that one can see digits 1, 2 and 3 (see the first, third and fifth subfigures of Fig.~\ref{fig:123_varying_results}).

\subsection{Circle-Square Dataset}
\label{sec:circle_sq}
\begin{figure}[H]
\includegraphics[width=0.49\linewidth]{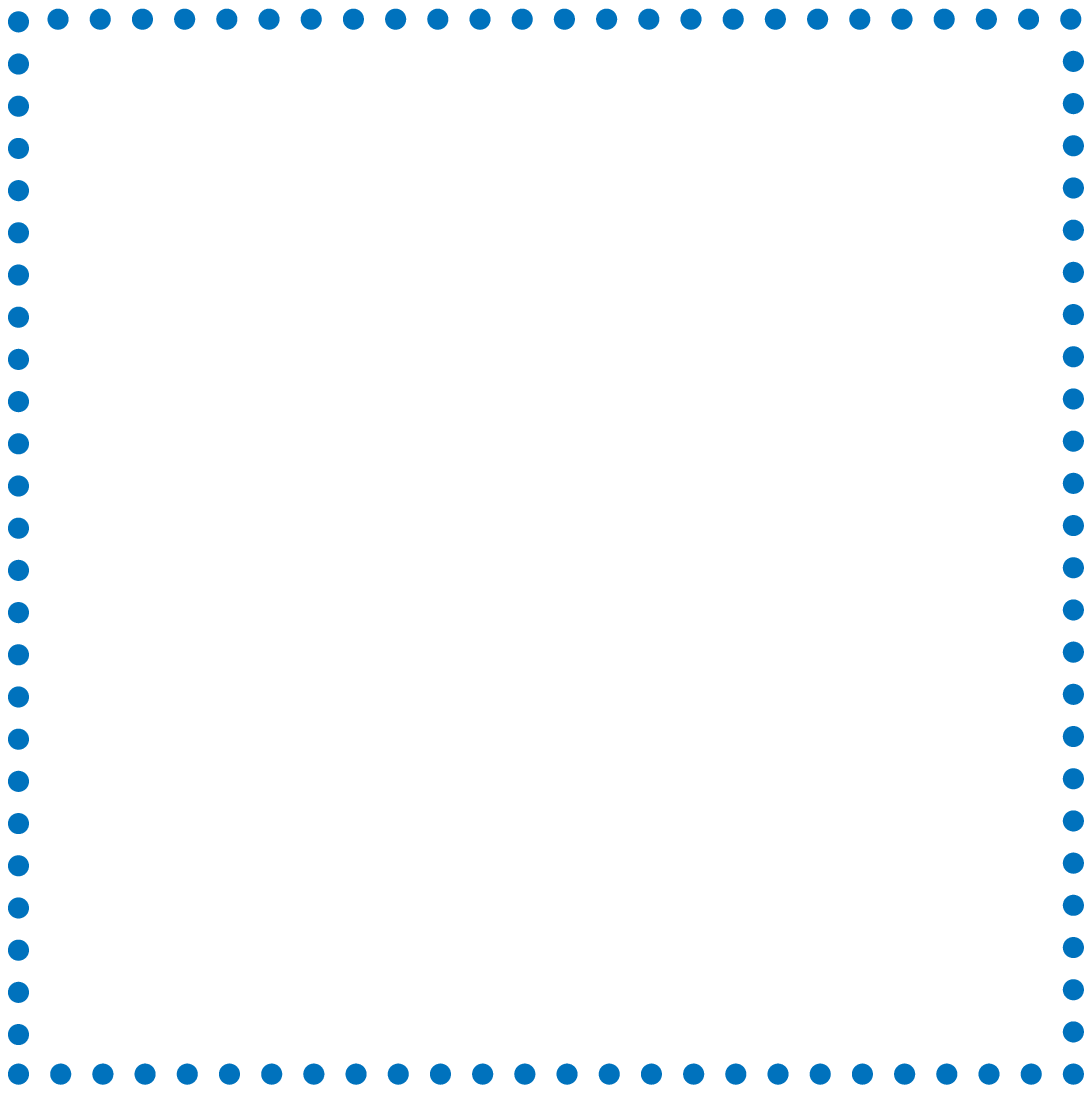}
\includegraphics[width=0.49\linewidth]{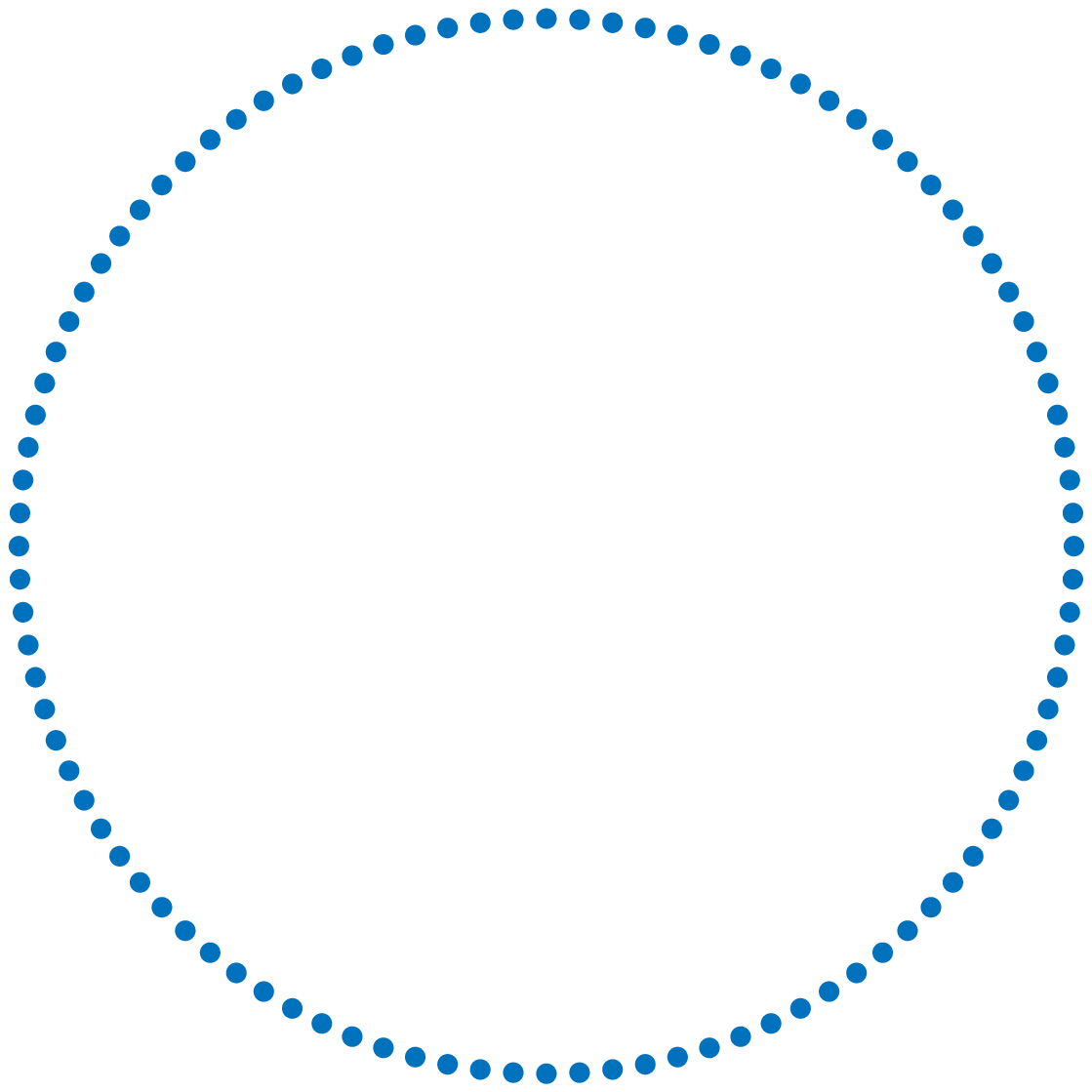}
\caption{Circle-Square dataset: the input is 100 points sampled from a square and a circle.}
\label{fig:circle_square_original}
\end{figure}

\begin{figure}[h]
\includegraphics[width=0.24\linewidth]{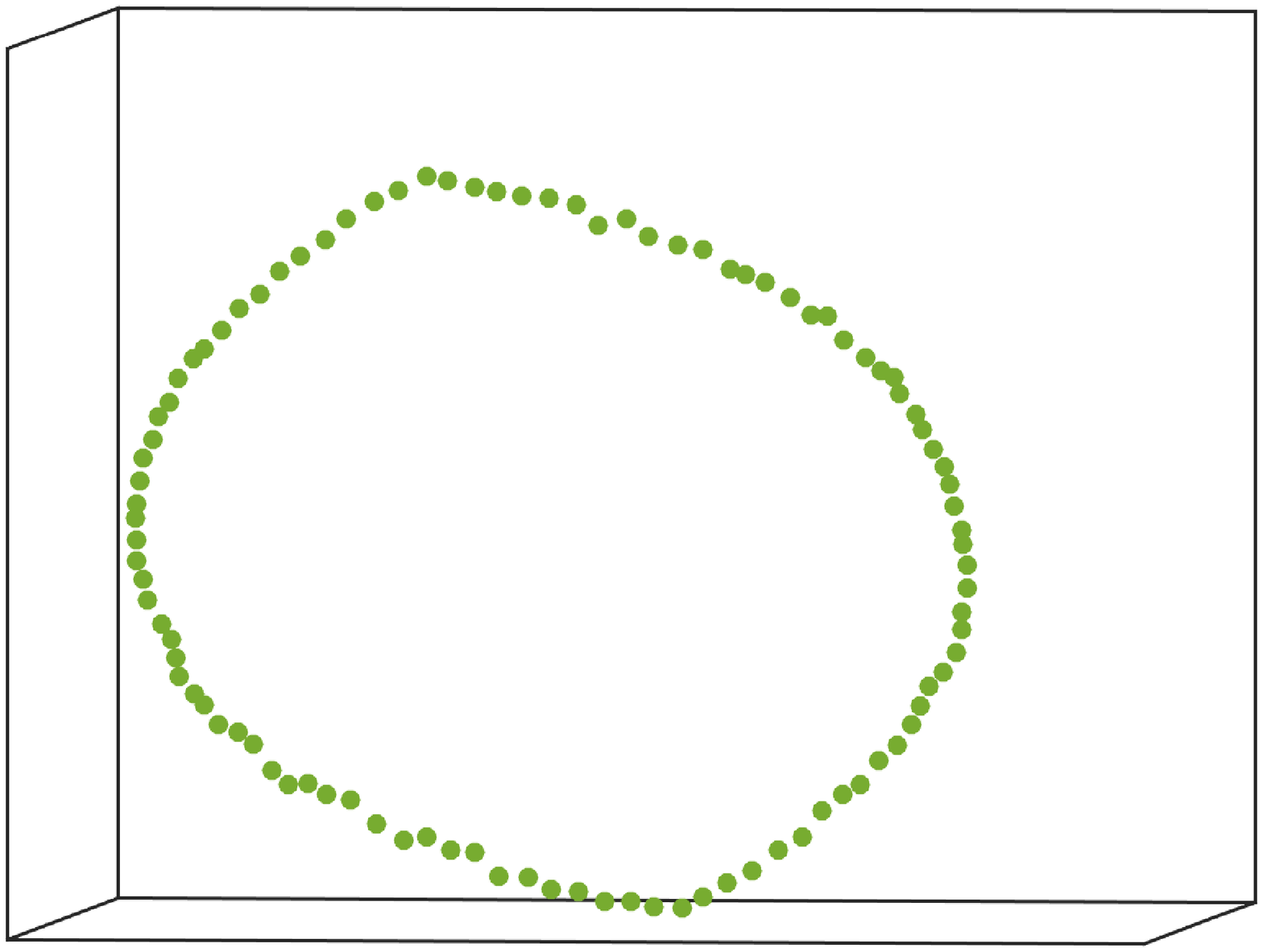}
\includegraphics[width=0.24\linewidth]{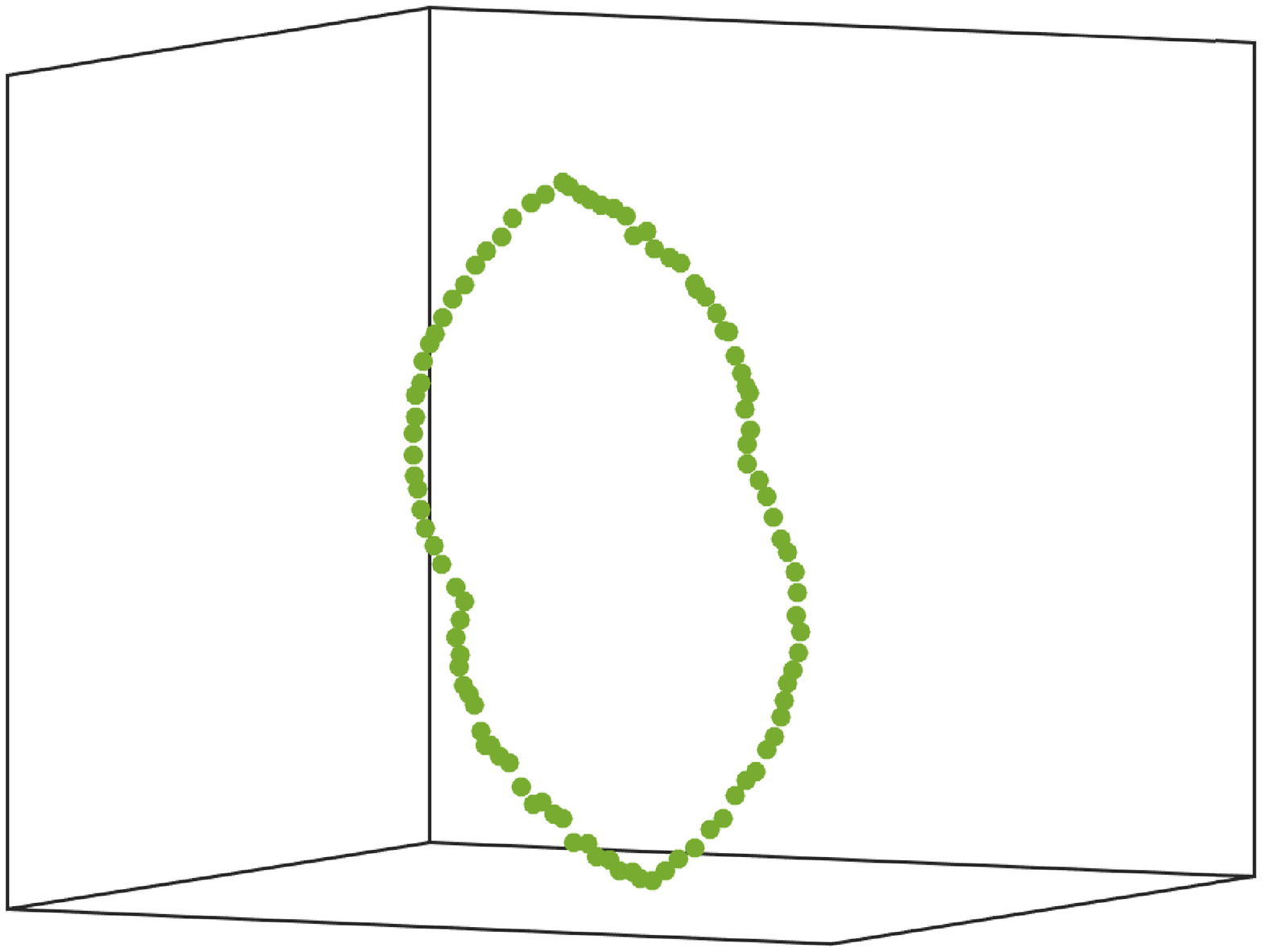}
\includegraphics[width=0.24\linewidth]{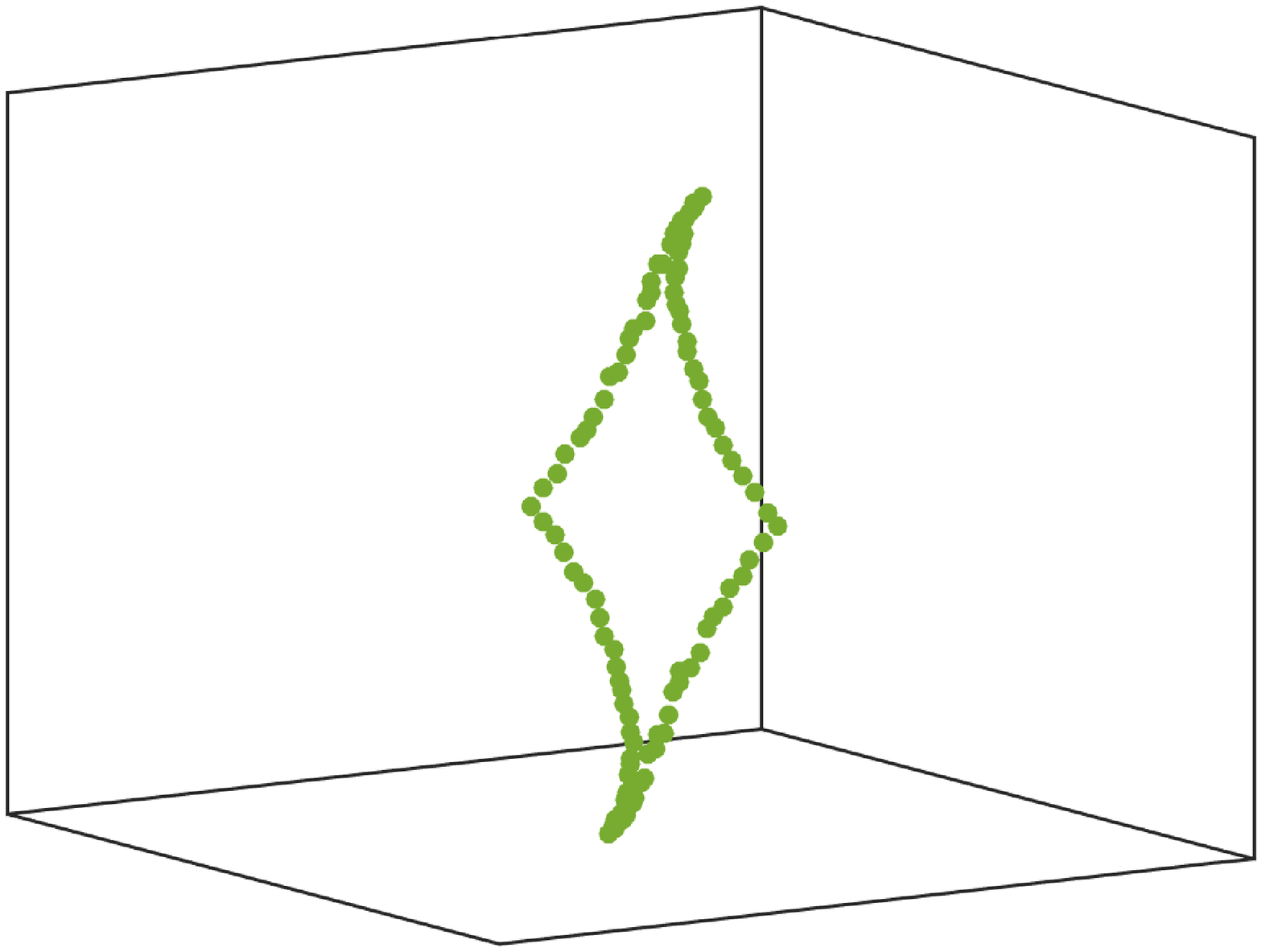}
\includegraphics[width=0.24\linewidth]{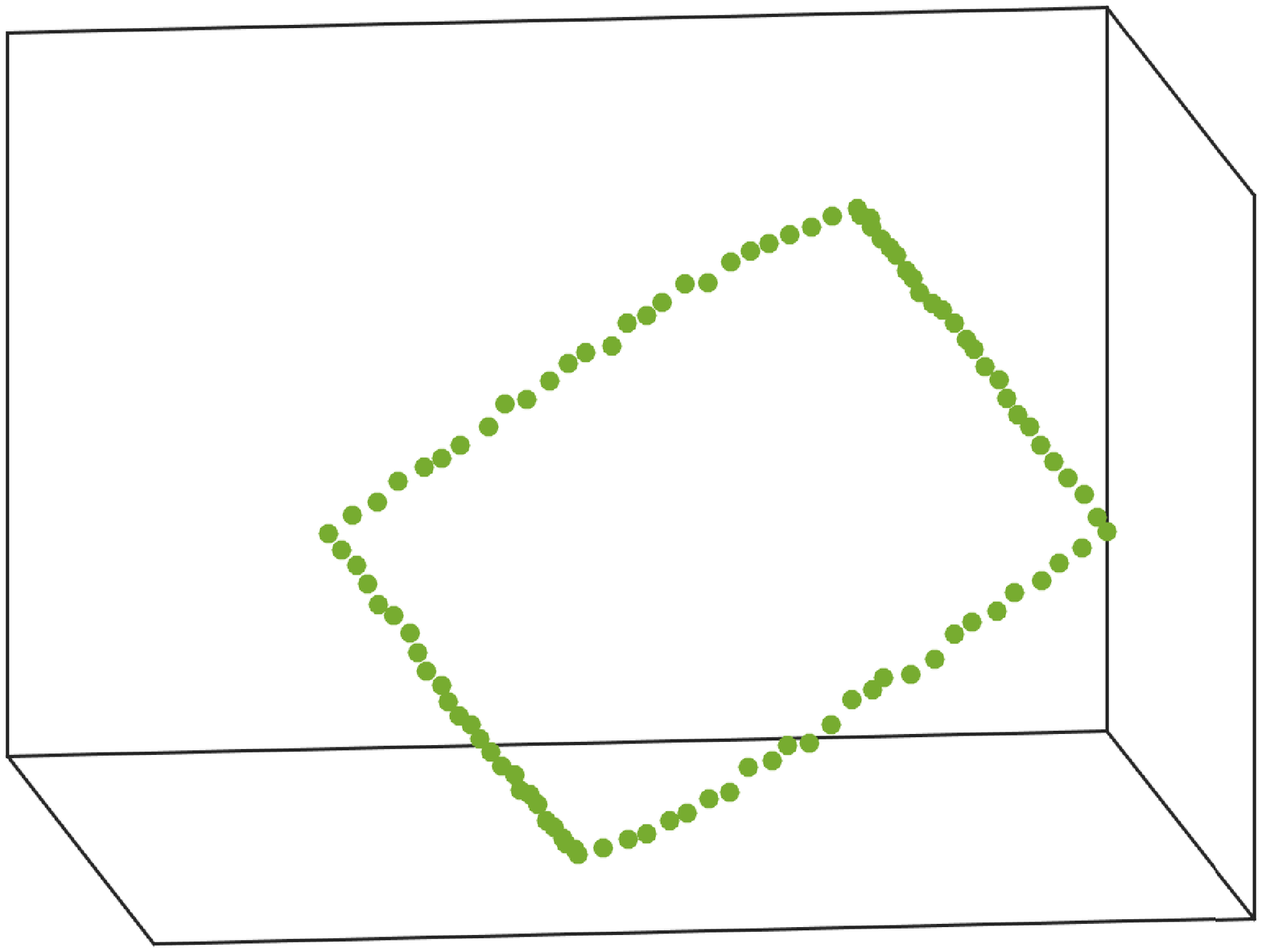}

\includegraphics[width=0.24\linewidth]{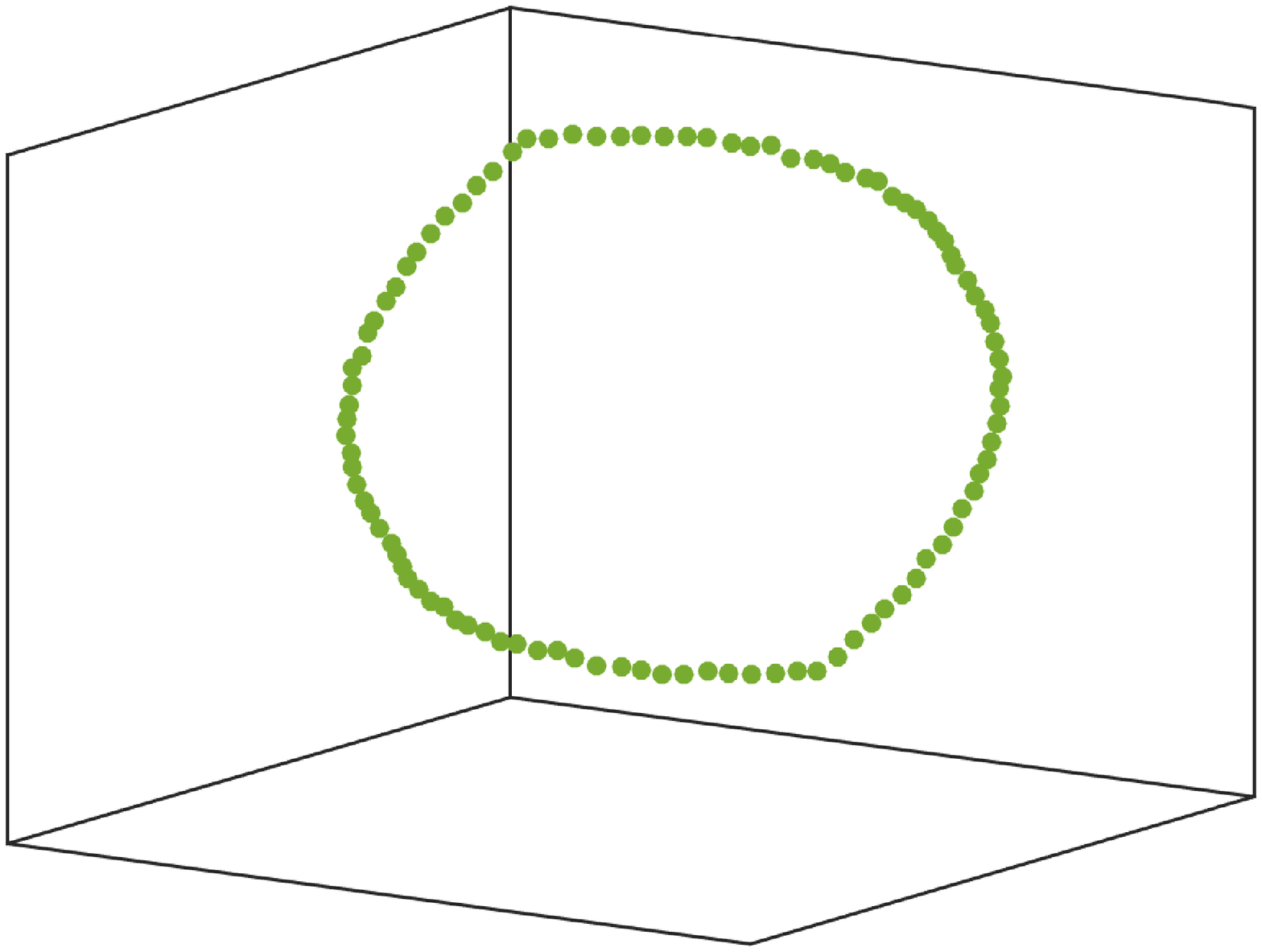}
\includegraphics[width=0.24\linewidth]{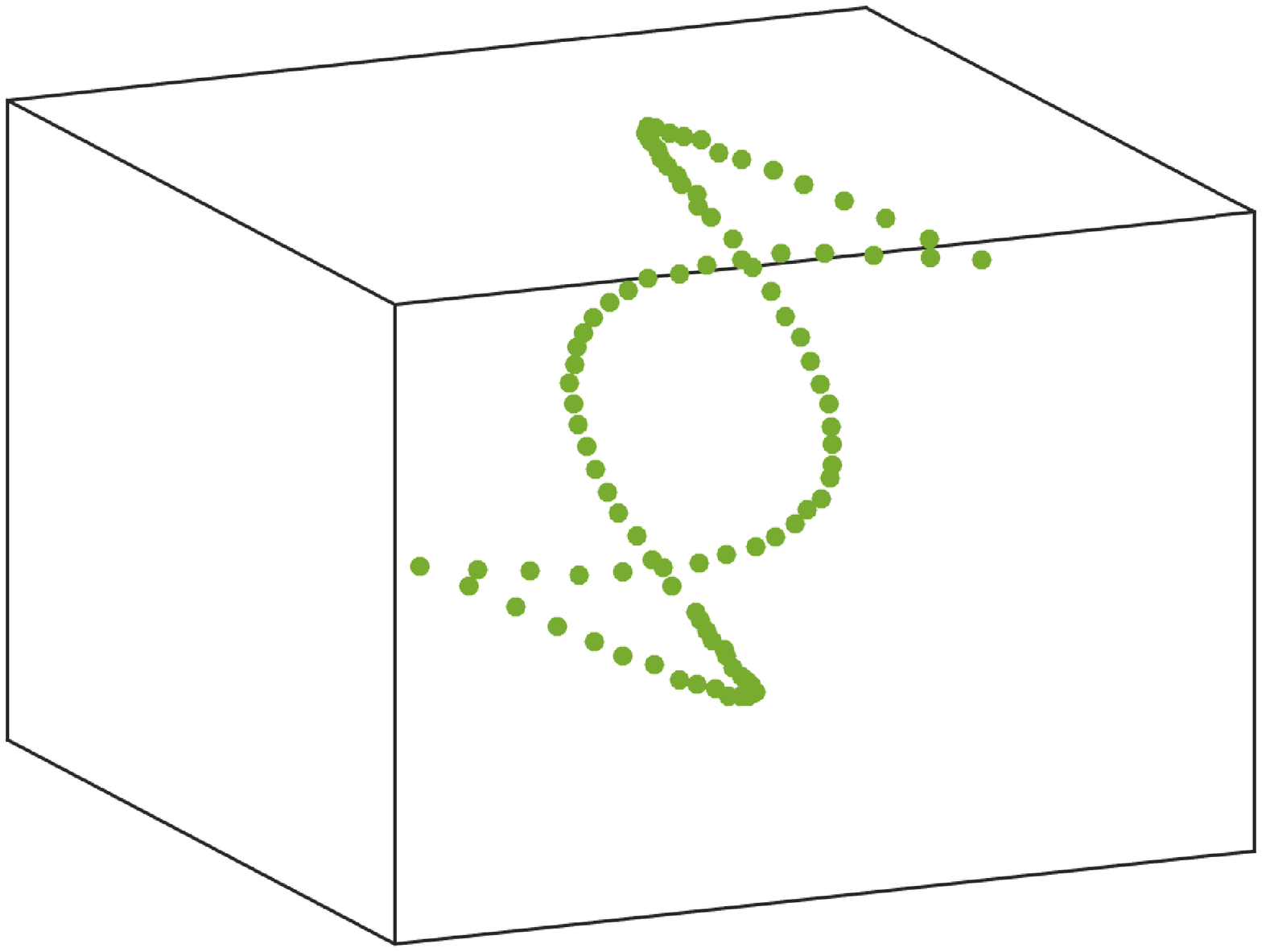}
\includegraphics[width=0.24\linewidth]{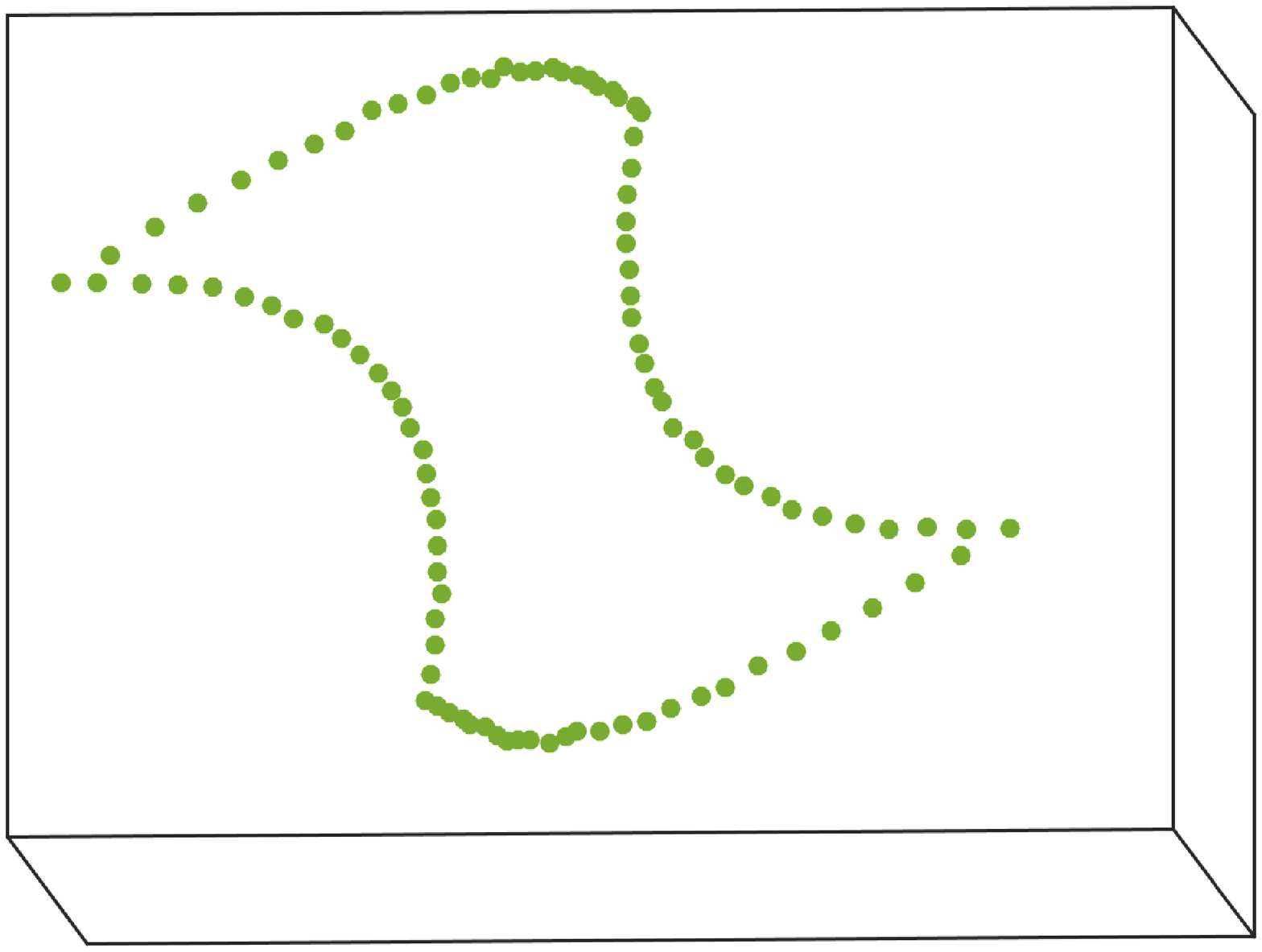}
\includegraphics[width=0.24\linewidth]{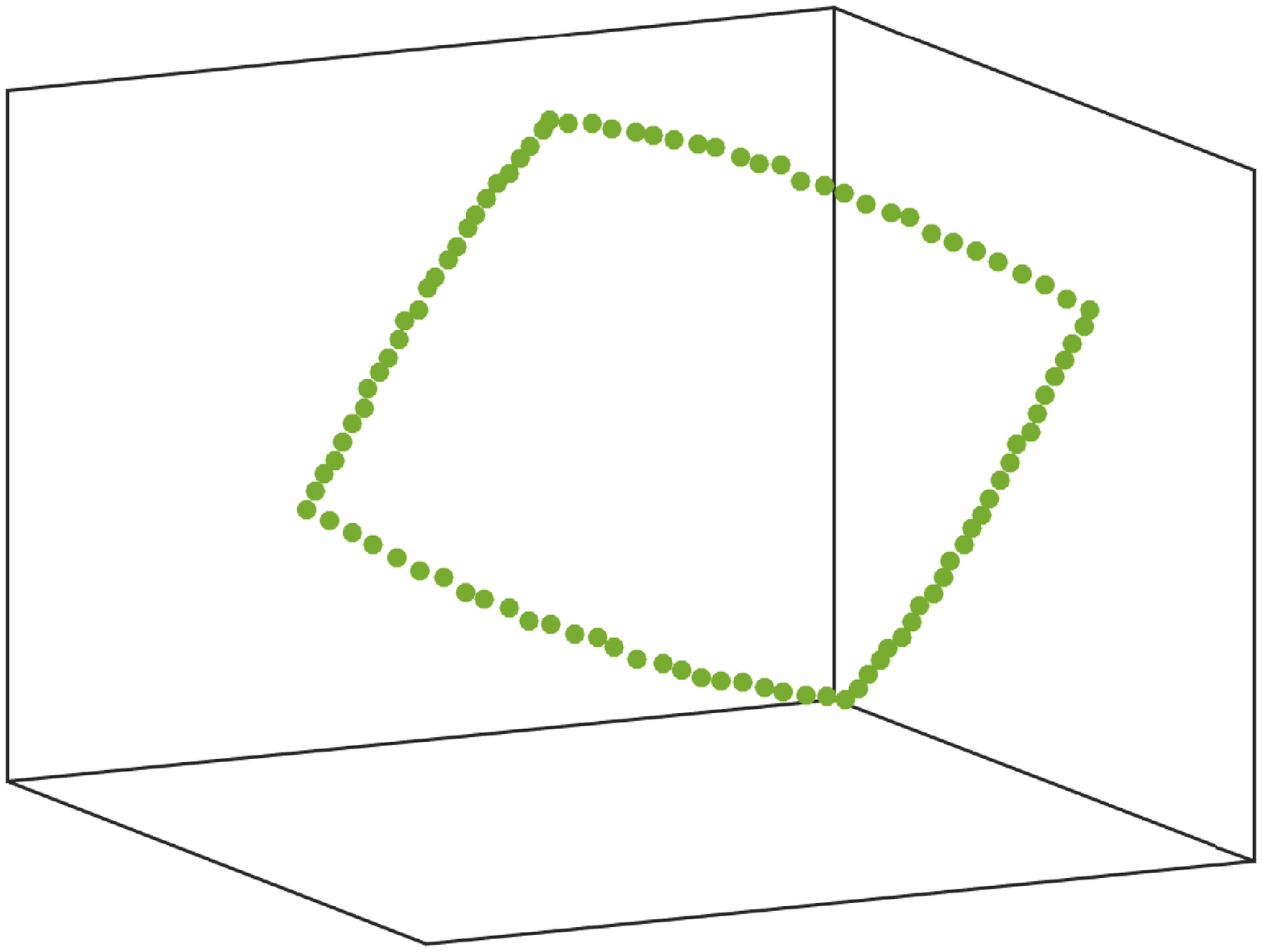}
\caption{The output of Algorithms~\ref{algo:fixed} and \ref{algo:varying} applied on the Circle-Square dataset. The first row show the result of Algorithm~\ref{algo:fixed} and the second row shows the result of Algorithm~\ref{algo:varying}. For both rows, the first column captures the recovered circle and the last row captures the recovered square. The second and third rows show the rotational transition in 3D between the circle and the square, revealing its 3D structure. The final MPSE stress for the circle-square dataset with fixed projections (Algorithm~\ref{algo:fixed})is 0.046, with projection-wise stress values of 0.045 and 0.047. The final MPSE stress for the circle-square dataset with varying projections (Algorithm~\ref{algo:varying}) is 0.044, with projection-wise stress values of 0.044 and 0.045.}
\label{fig:circle_squire_all_results}
\end{figure}

The Circle-Square dataset is also an example of 3D share reconstruction from 2D. For this dataset we construct 2D images of a unit square and a circle with radius 1. From these figures we sample 100 points; see Fig.~\ref{fig:circle_square_original}. We create the corresponding two distance matrices and run the MPSE algorithms.
We first run Algorithm~\ref{algo:fixed} for the distance matrices created from the Circle-Square dataset; see Fig.~\ref{fig:circle_square_original}. For the algorithm we used the following parameters: maximum number of iterations $T = 500$, fixed projections $\mP^1$, and $\mP^2$ from \eqref{eq:60deg_proj},  starting learning rate $\mu_0 = 1$ and stochastic constant $c = 0.01$ with random initialization. The results are shown in the first row of Fig.~\ref{fig:circle_squire_all_results}. The algorithm successfully placed the points in 3D so that one can see a circle and a square from different planes (first and last subfigures).
The second and third subfigures show the rotational transition in 3D between the circle and the square.

\begin{figure*}
\includegraphics[width=0.24\linewidth]{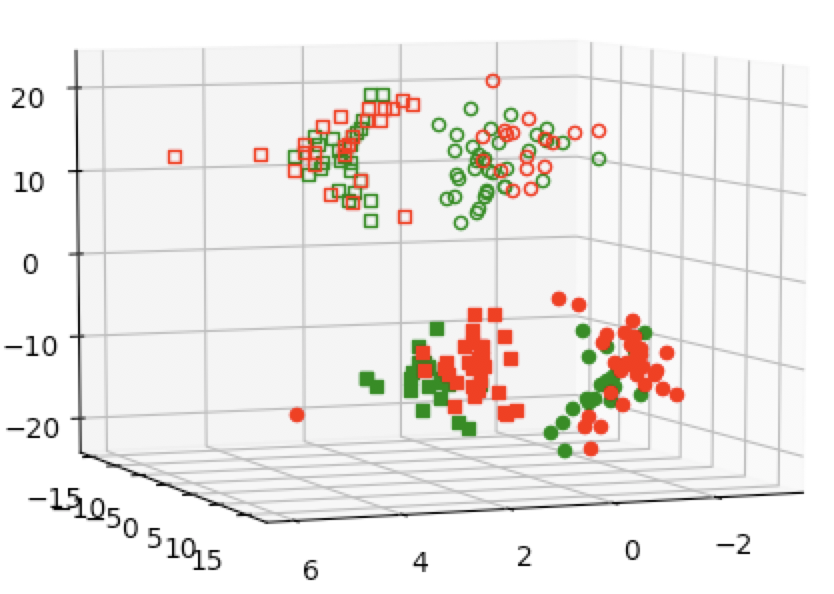}
\includegraphics[width=0.24\linewidth]{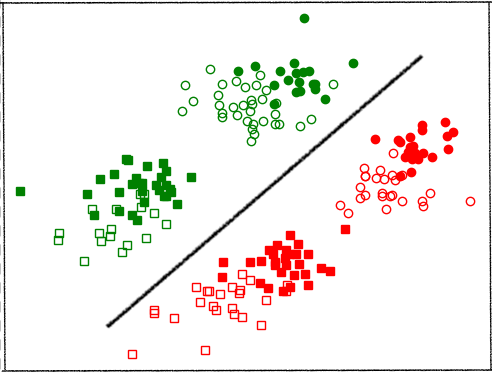}
\includegraphics[width=0.24\linewidth]{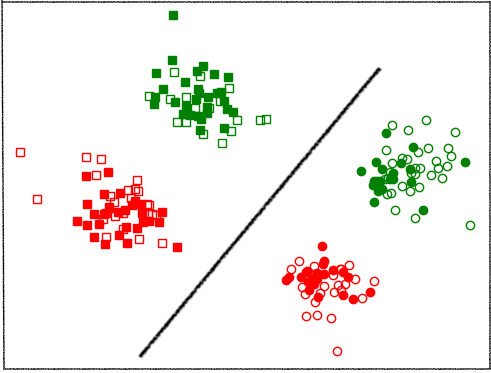}
\includegraphics[width=0.24\linewidth]{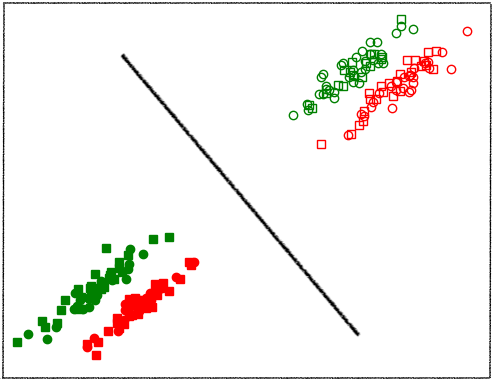}
\caption{The output of Algorithm~\ref{algo:varying} on the Clustering dataset. The first subfigure shows the 3D embedding and the remaining ones correspond to the 3 projections. To make it easier to see the 3 different types of clusters captured here, in all subfigures we use green/red to distinguish between datapoints from different clusters of the first type, circle/square glyphs for clusters of the second type, and filled/empty glyphs for clusters of the third type. We manually add the separating lines to highlight the separations realized in the 3 planes. The final MPSE  stress  for  the  Clustering dataset with varying projections (Algorithm~\ref{algo:varying}) is 0.60, with projection-wise stress values of 0.52, 0.47 and 0.77.}
\label{fig:cluster_data_results}
\end{figure*}

Next, we run Algorithm~\ref{algo:varying} for the same Circle-Square dataset with the following parameters: maximum number of iterations $T = 100$, the starting learning rate $\mu_0 = 1$ and stochastic constant $c = 0.01$ with smart initialization described in Algorithm~\ref{algo:initialization}. The results are shown in the second row of Fig.~\ref{fig:circle_squire_all_results}. As we can see the algorithm again successfully placed the points in 3D so that one can see a circle and a square from different directions (first and last subfigures).

\subsection{Clusters Dataset}
One of the many applications of dimensionality reduction is to preprocess the dataset by reducing its dimension and then apply a clustering/classification algorithm. The setting where we want to test whether our proposed algorithm preserves the clusters of a given dataset is the following: We want to visualize a dataset in 3D such that different 2D projections capture different clusters (say, determined by considering different attributes of the data). For this purpose, we fix $n = 200$ and create 3 datasets in 2D each having 2 well separated clusters of 100 datapoints each. 
\begin{figure}[h]
 \includegraphics[width=0.32\linewidth]{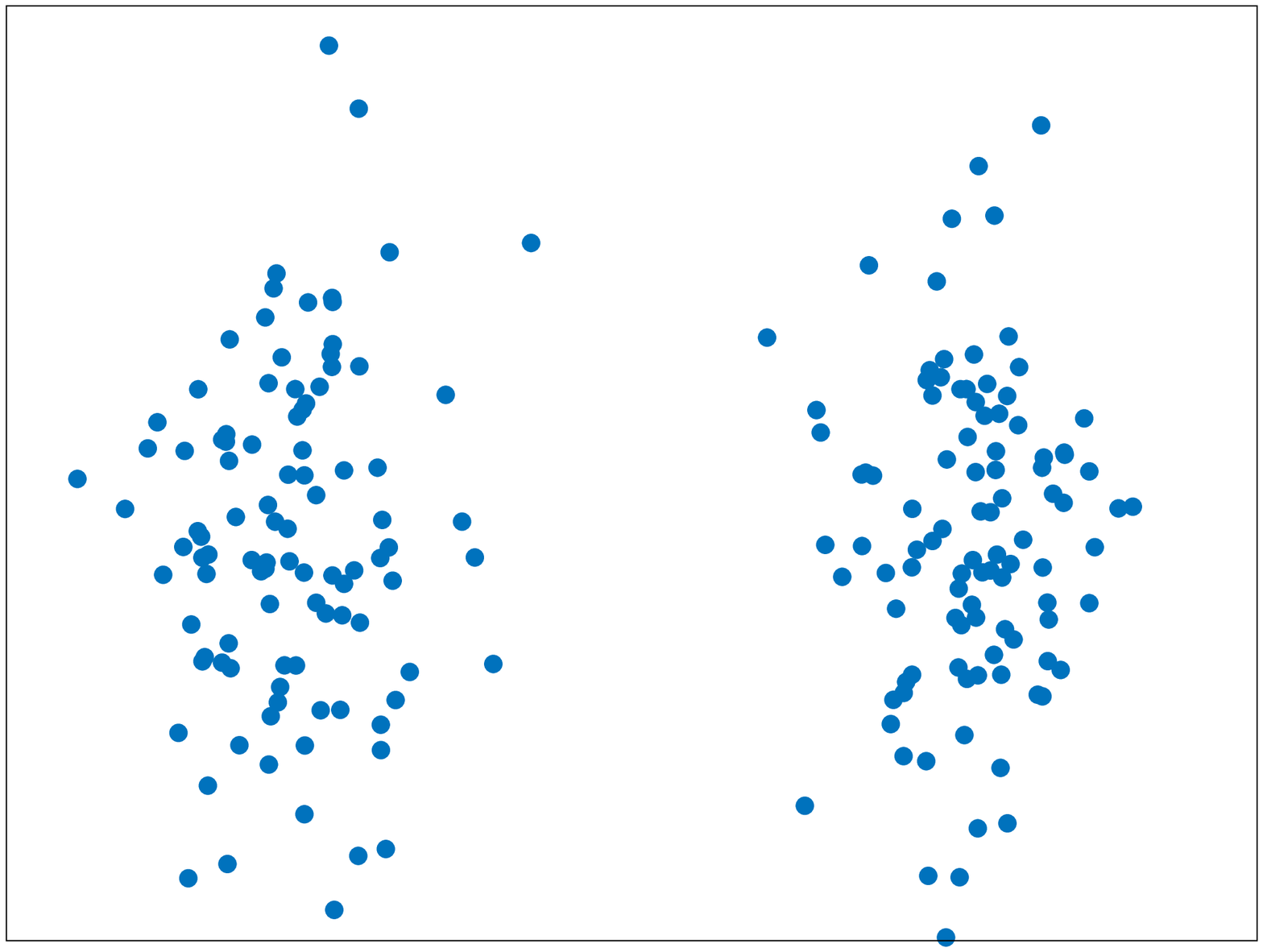}
 \includegraphics[width=0.32\linewidth]{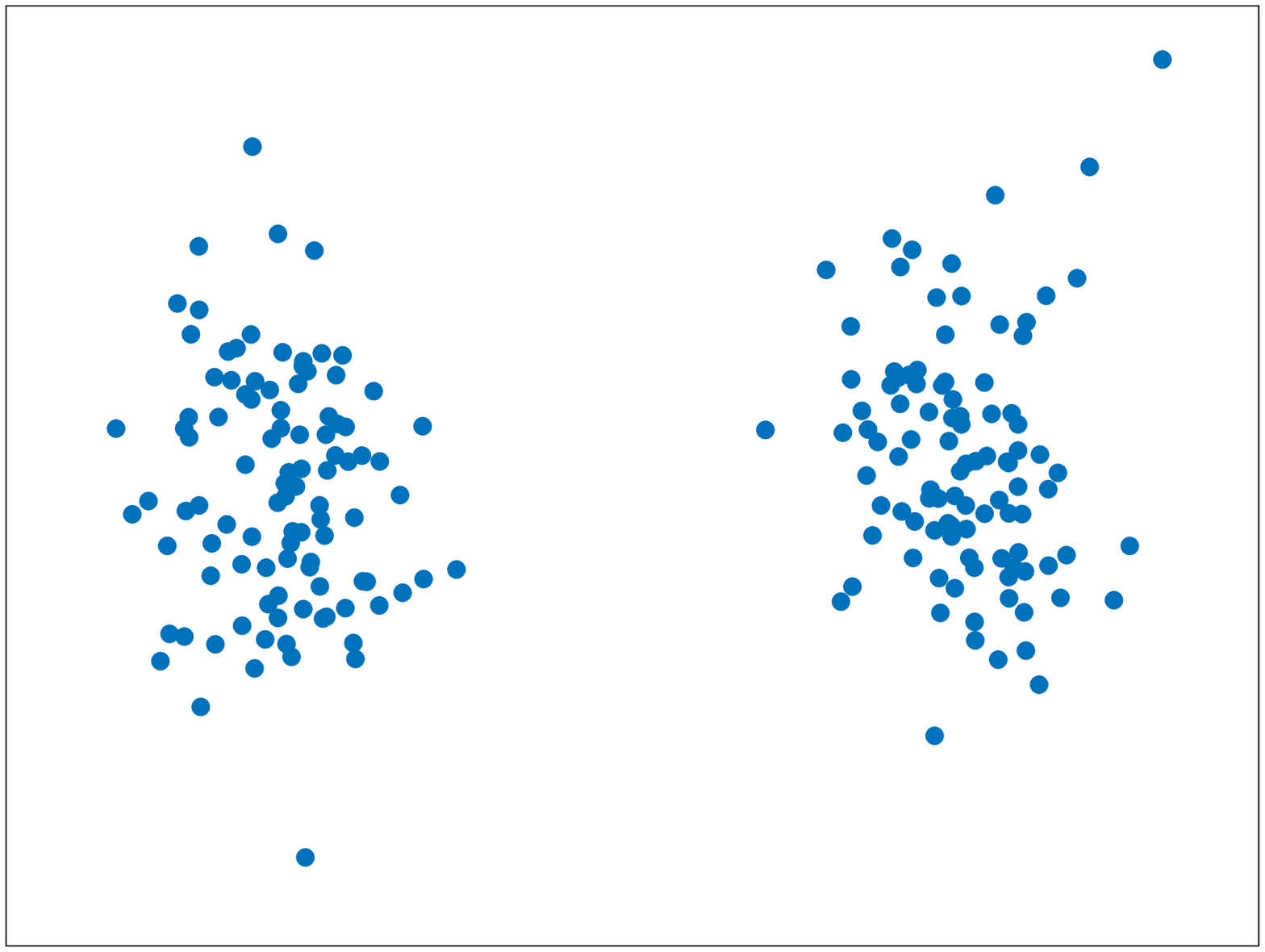}
 \includegraphics[width=0.32\linewidth]{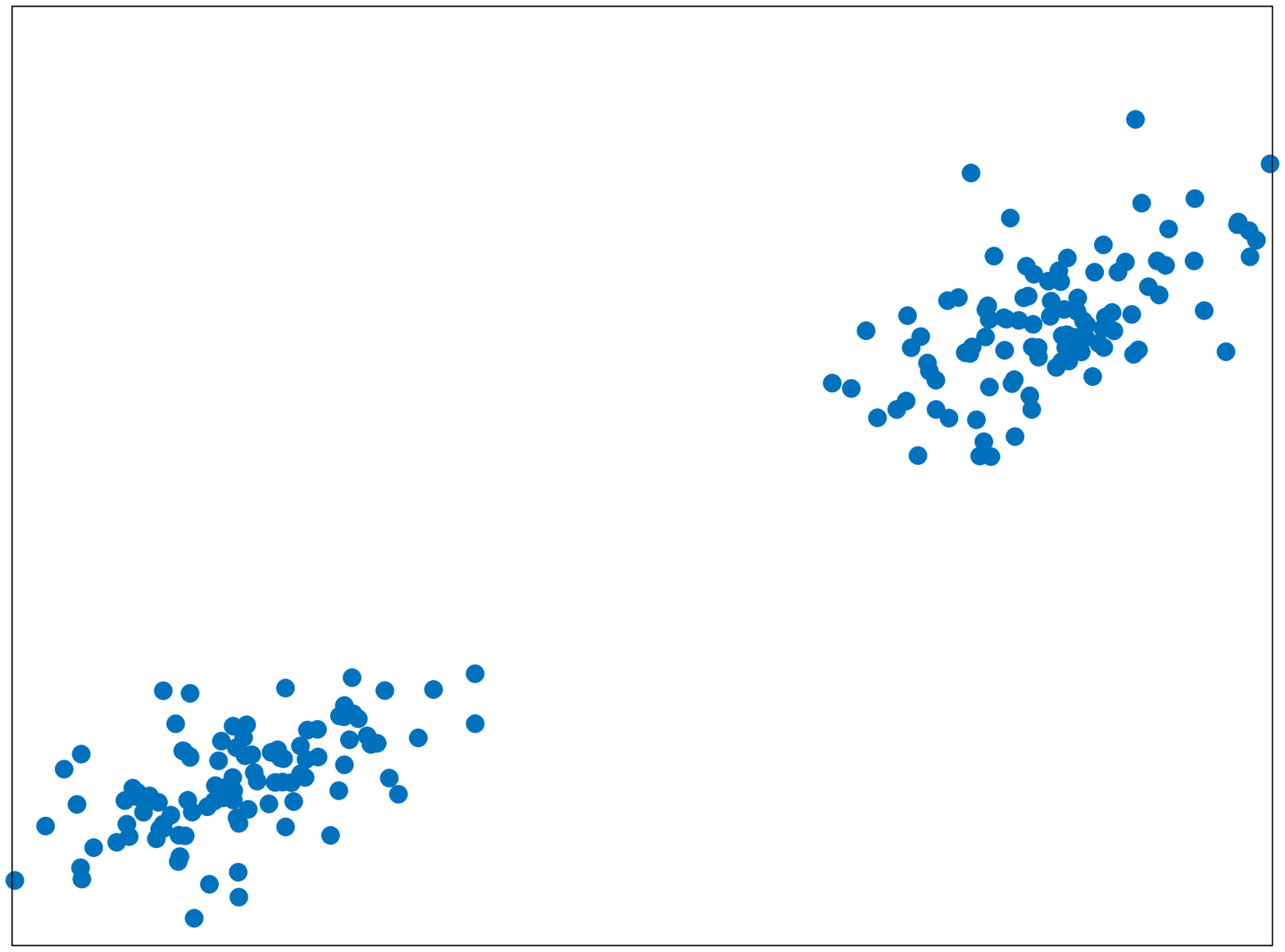}
 \caption{The input to the Clusters dataset: each subfigure shows an example of a dataset of 200 points with 2 well separated clusters of 100 points.}
\label{fig:cluster_data_init}
\end{figure}
Our goal is to apply Algorithms~\ref{algo:fixed} and \ref{algo:varying} and check whether the corresponding 3D embeddings preserve the clusters and if so what the corresponding projections look like. We apply Algorithm~\ref{algo:varying} with $T = 100$, the distance matrices corresponding to datapoints of subfigures of Fig.~\ref{fig:cluster_data_init} and random initialization. The results are shown in Fig.~\ref{fig:cluster_data_results}. 

Remarkably, as shown in Fig.~\ref{fig:cluster_data_results}, the MPSE algorithm with varying projections captured the special structure of the dataset (three different pairs of clusters defined on the same datapoints) and embedded them in 3D so that the 3 different separations can be seen from the corresponding projections.

\subsection{15th-century Florentine Families Dataset}
\label{sec:florence}

We now consider a network dataset that captures social relationships between prominent families in  Renaissance Florence (one of the largest and richest cities in 15th century Europe)  \cite{Padgett1993florence,Sharfie2015multigraph}. This dataset contains descriptions of social ties between Florentine families during the years 1426-1434, a period of historical significance that marks the rise to power of the Medici family.

The social ties are divided into categories such as `marriage' (number of marriages between two families) and `loan' (number of loans between two families). We can interpret this dataset as a network with multiple attributes. To visualize this network, we select a subset of the families such that the network structure for the `marriage' (1) and `loan' (2) attributes each form a connected graph. For each type of relation $k\in\{1,2\}$, we define the dissimilarity between two families as follows: for families $i$ and $j$, if there is a nonzero number $n^{(k)}_{ij}$ of ties between the two families, then the pairwise dissimilarity is initially given by $D^{(k)}_{ij}=1/n^{(k)}_{ij}$; afterwards, shortest path distances are computed for all pairs of families using this initial set of dissimilarities, resulting in dissimilarities for every pair of families. The pairwise weights used to produce the MDS \eqref{eq:normalized-mds-stress} and MPSE \eqref{eq:normalized-mpse-stress} embeddings are defined by $w^{(k)}_{ij}=1/D^{(k)}_{ij}$.

We compute a 2D MDS embedding based on marriage distances and a 2D MDS embedding based on loan distances (independent of each other). 
We then compute a 3D simultaneous embedding based on both marriage and loan distances using MPSE with variable projections with the following parameters: maximum number of iterations $T=300$, starting learning rate $\mu_0=1$, stochastic constant  $c=0.1$, and initial embedding from Algorithm~\ref{algo:initialization}.

\begin{figure*}
\includegraphics[width=0.245\linewidth]{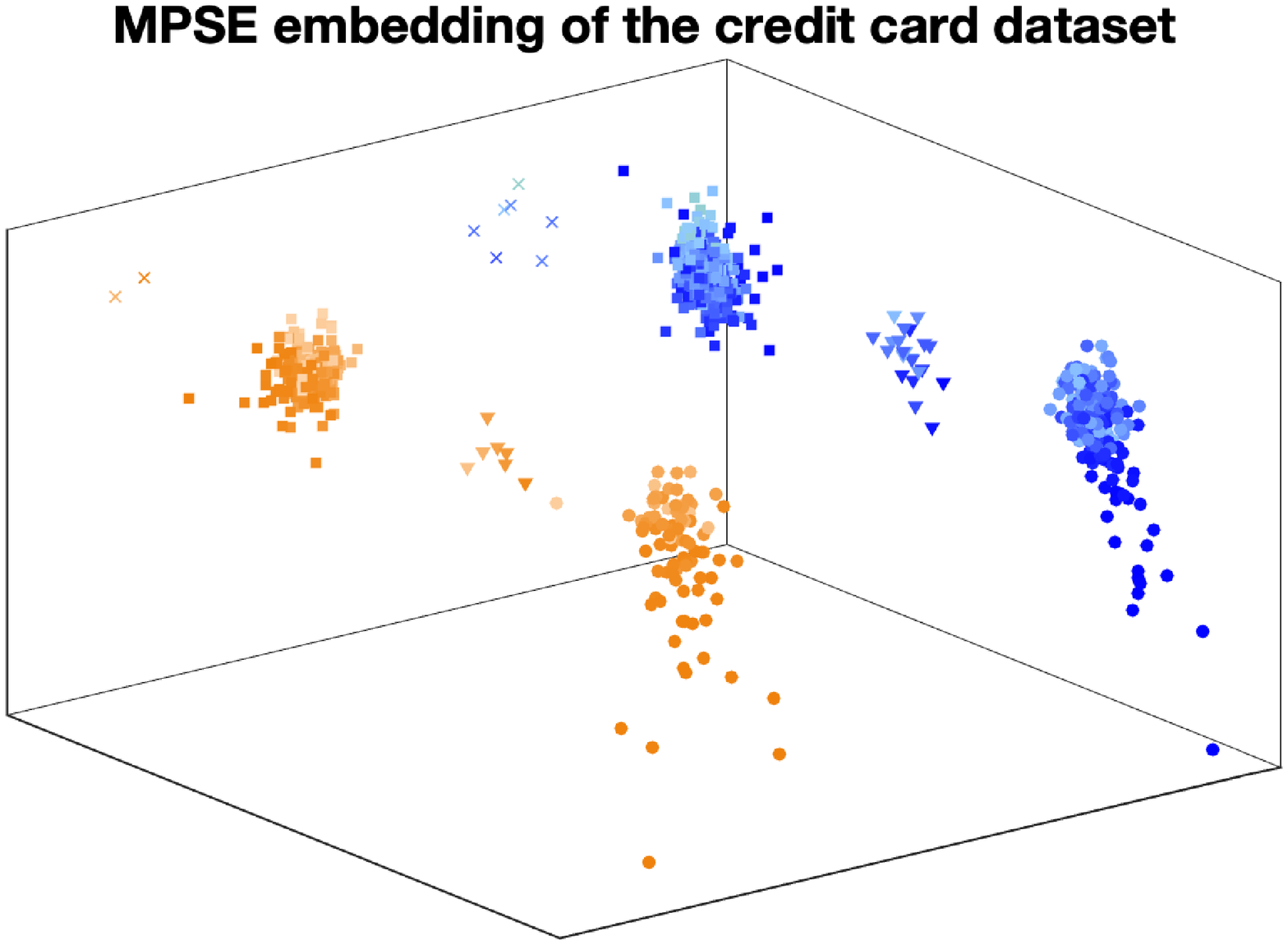}
\includegraphics[width=0.245\linewidth]{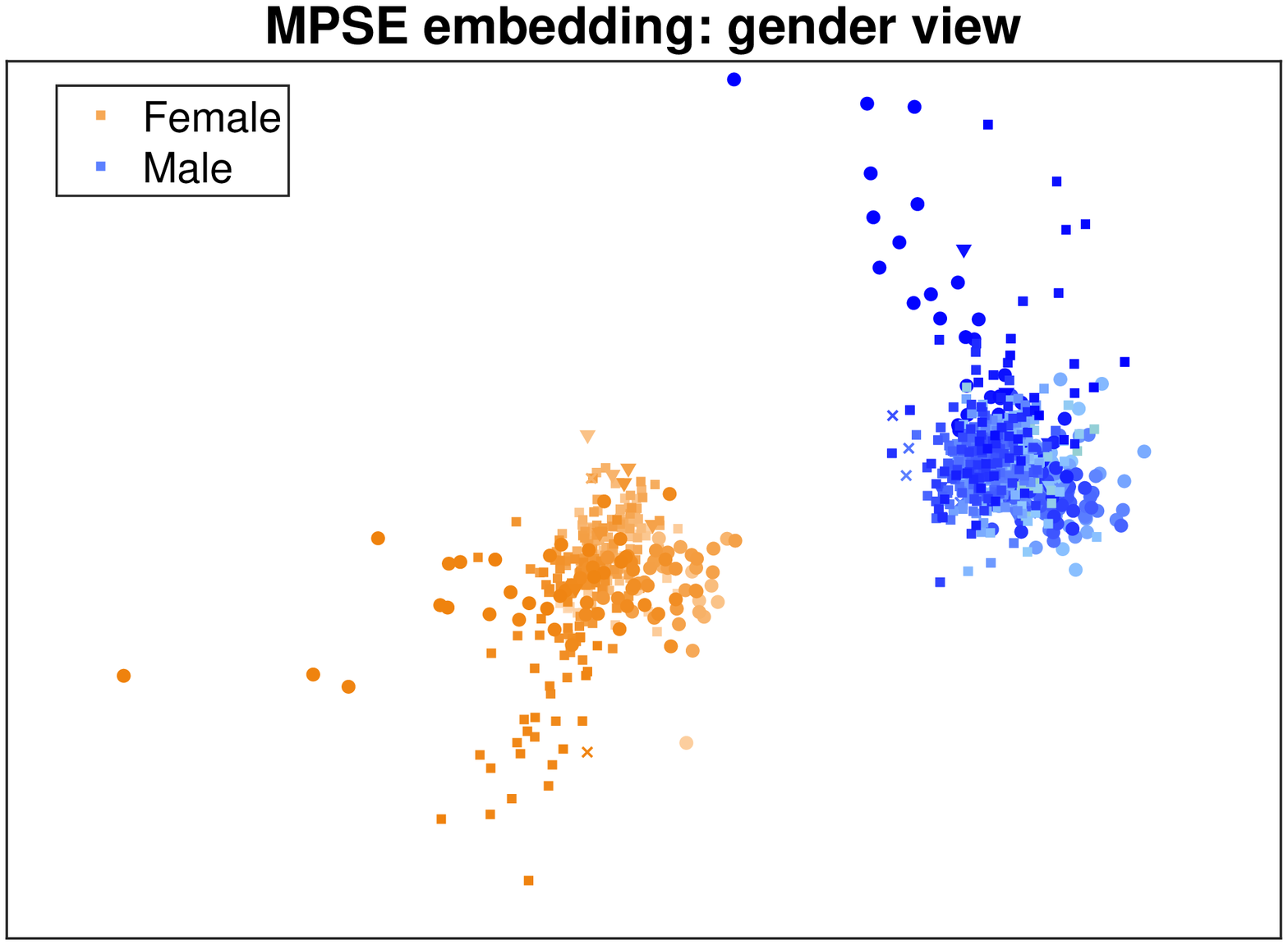}
\includegraphics[width=0.245\linewidth]{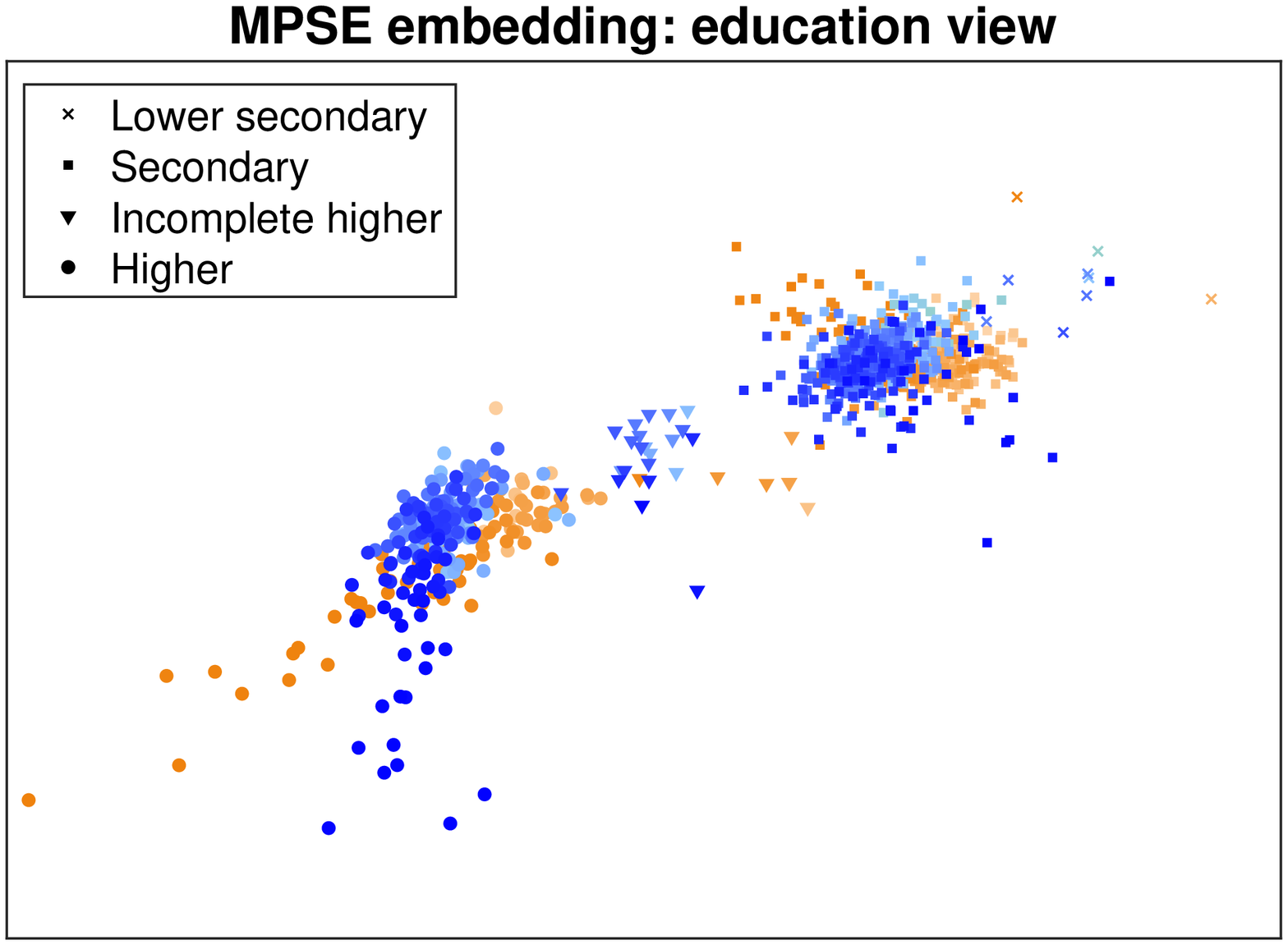}
\includegraphics[width=0.245\linewidth]{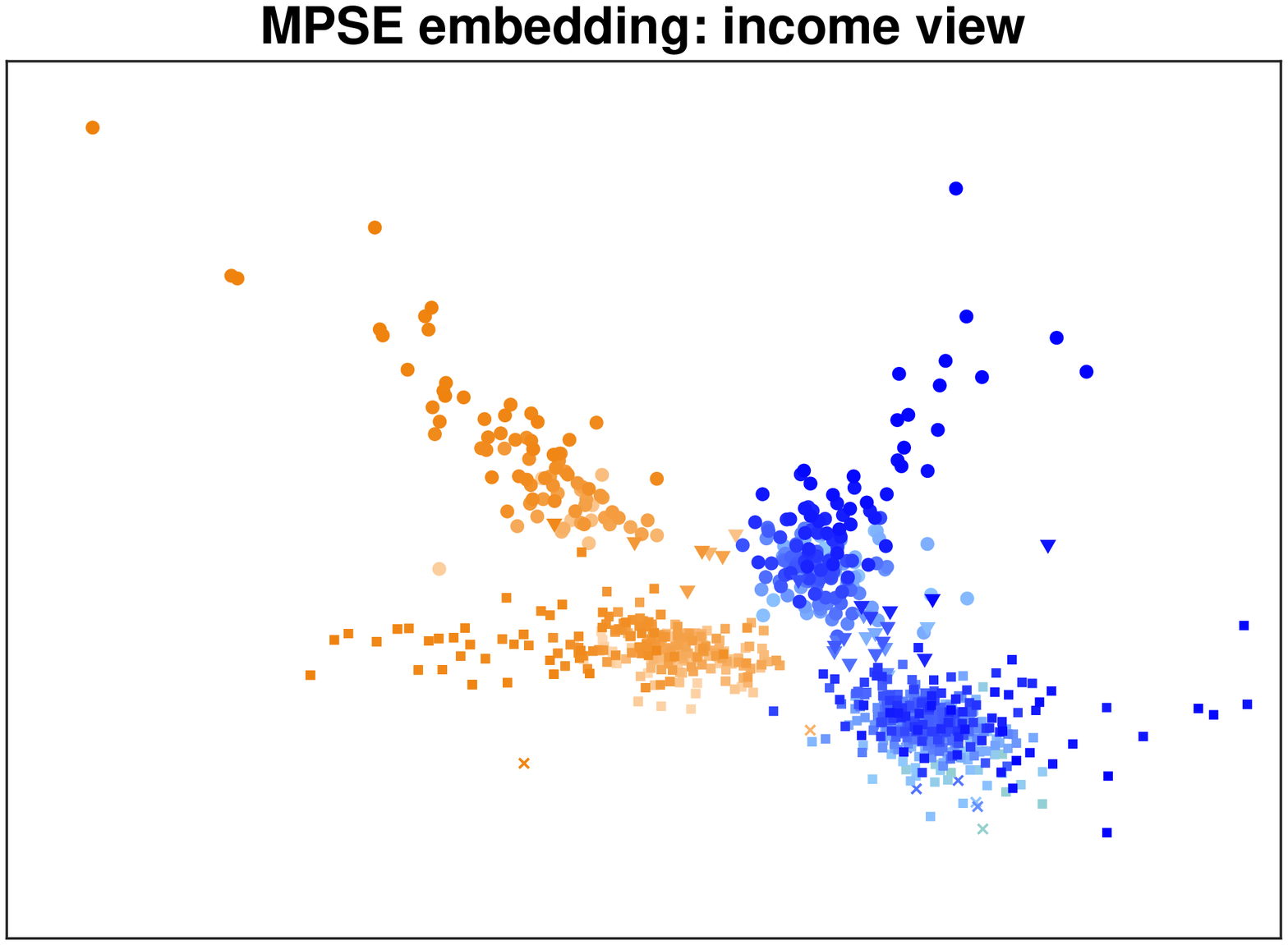}

\caption{The output of Algorithm~\ref{algo:varying} on the credit card dataset. The first subfigure shows the 3D embedding and the remaining ones correspond to the 3 projections on gender, education and income views, respectively. In all figures, high color intensity represents high income, color represents gender, and glyph shapes represent education levels. The final MPSE stress for the credit card dataset with varying projections is 0.40, with projection-wise stress values of 0.28, 0.29 and 0.55.}
\label{fig:ccapp}
\end{figure*}

\begin{figure}[h]
\includegraphics[width=0.49\linewidth]{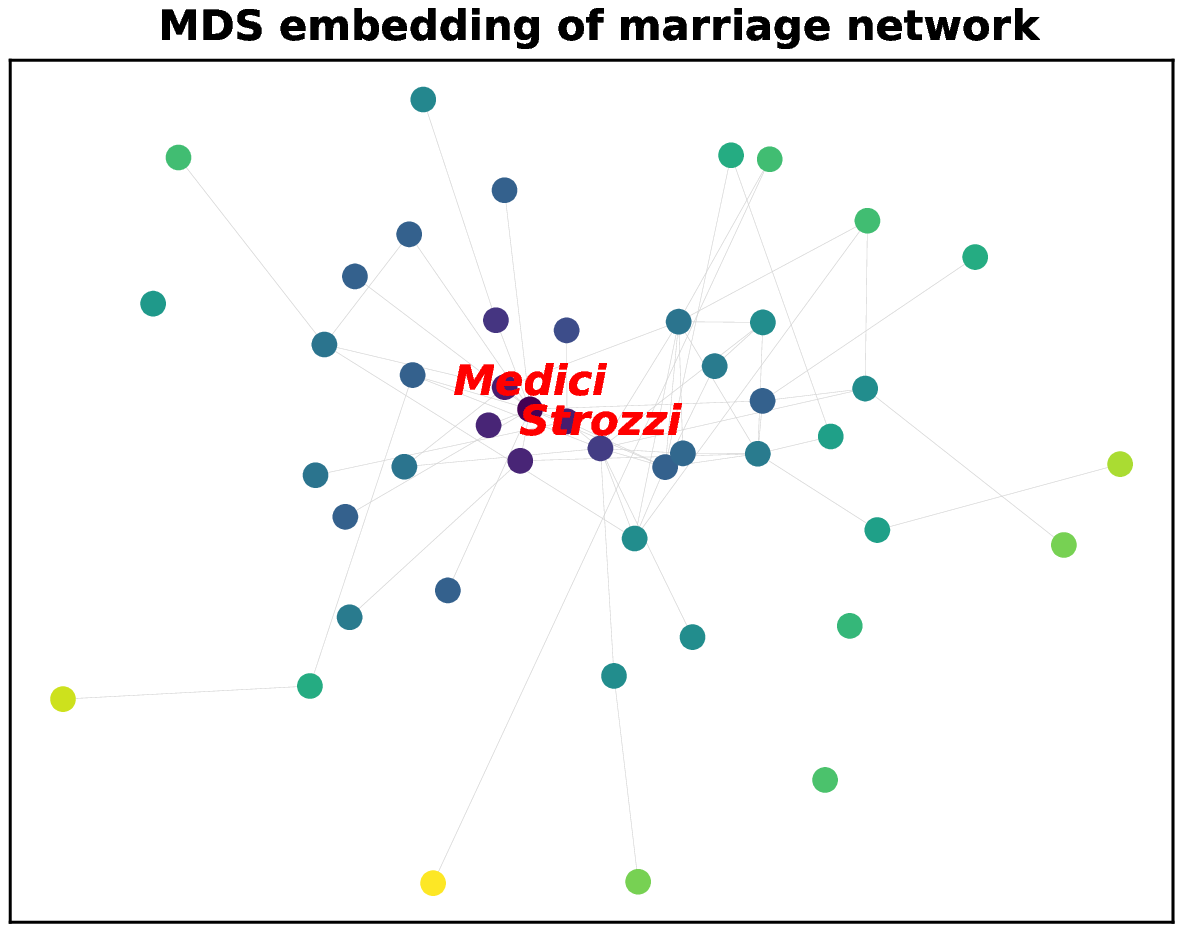}
\includegraphics[width=0.49\linewidth]{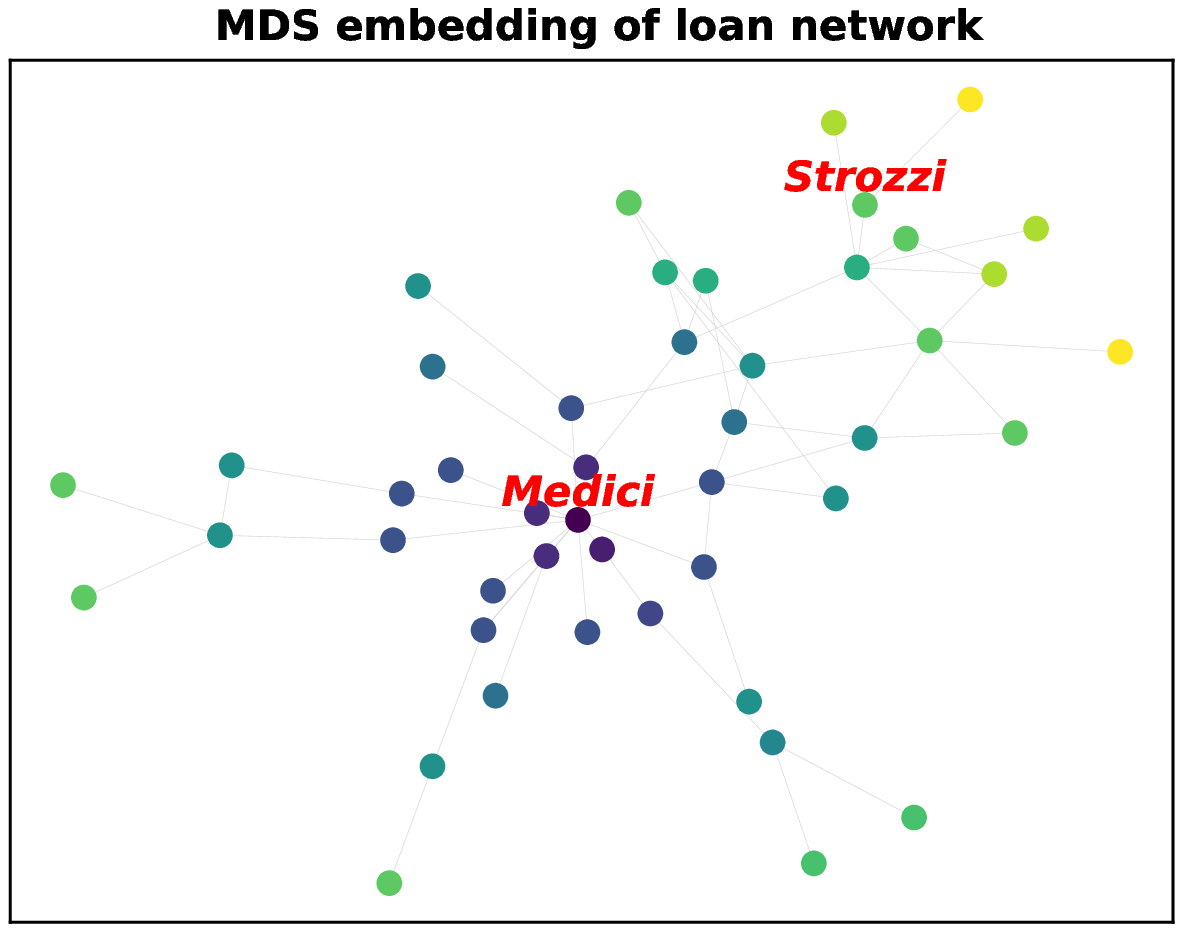}
\includegraphics[width=0.49\linewidth]{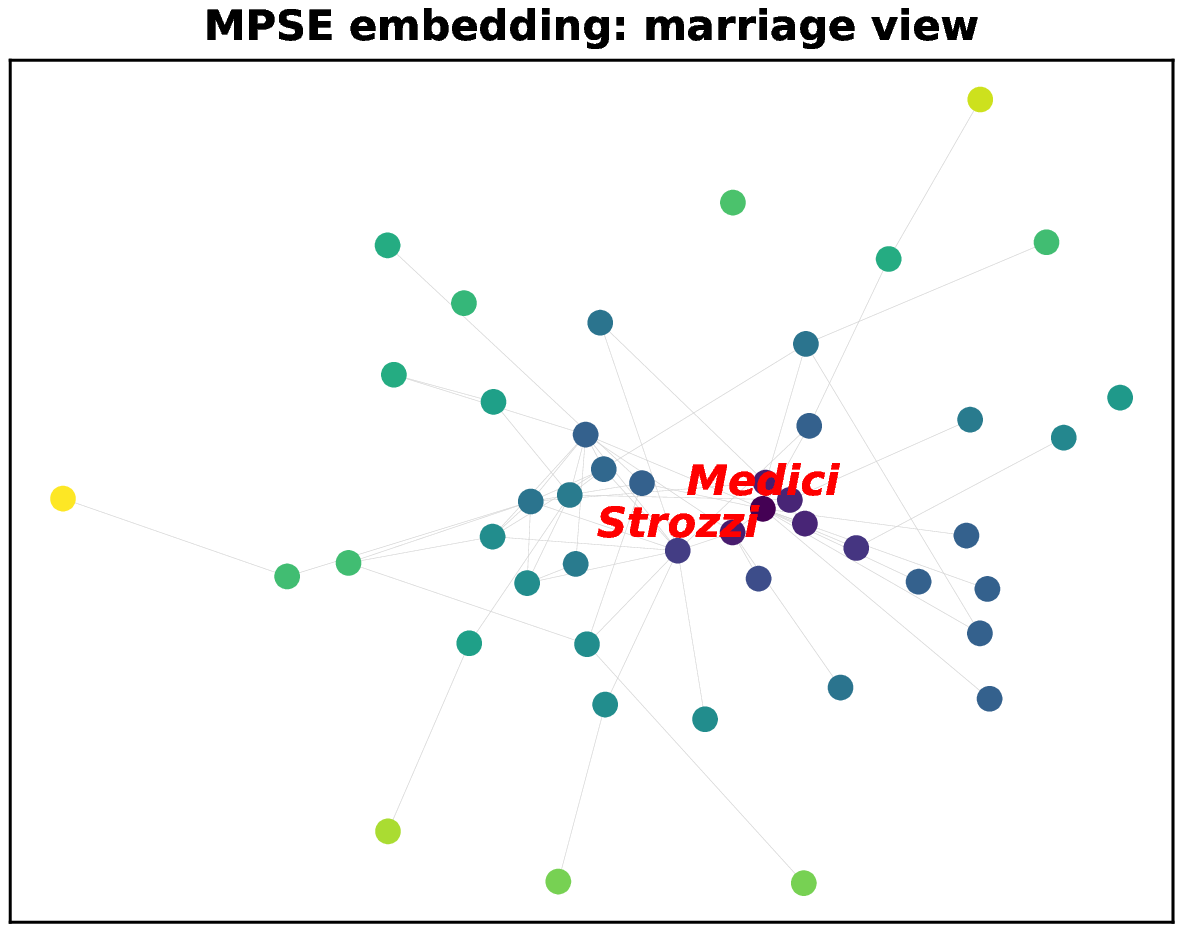}
\includegraphics[width=0.49\linewidth]{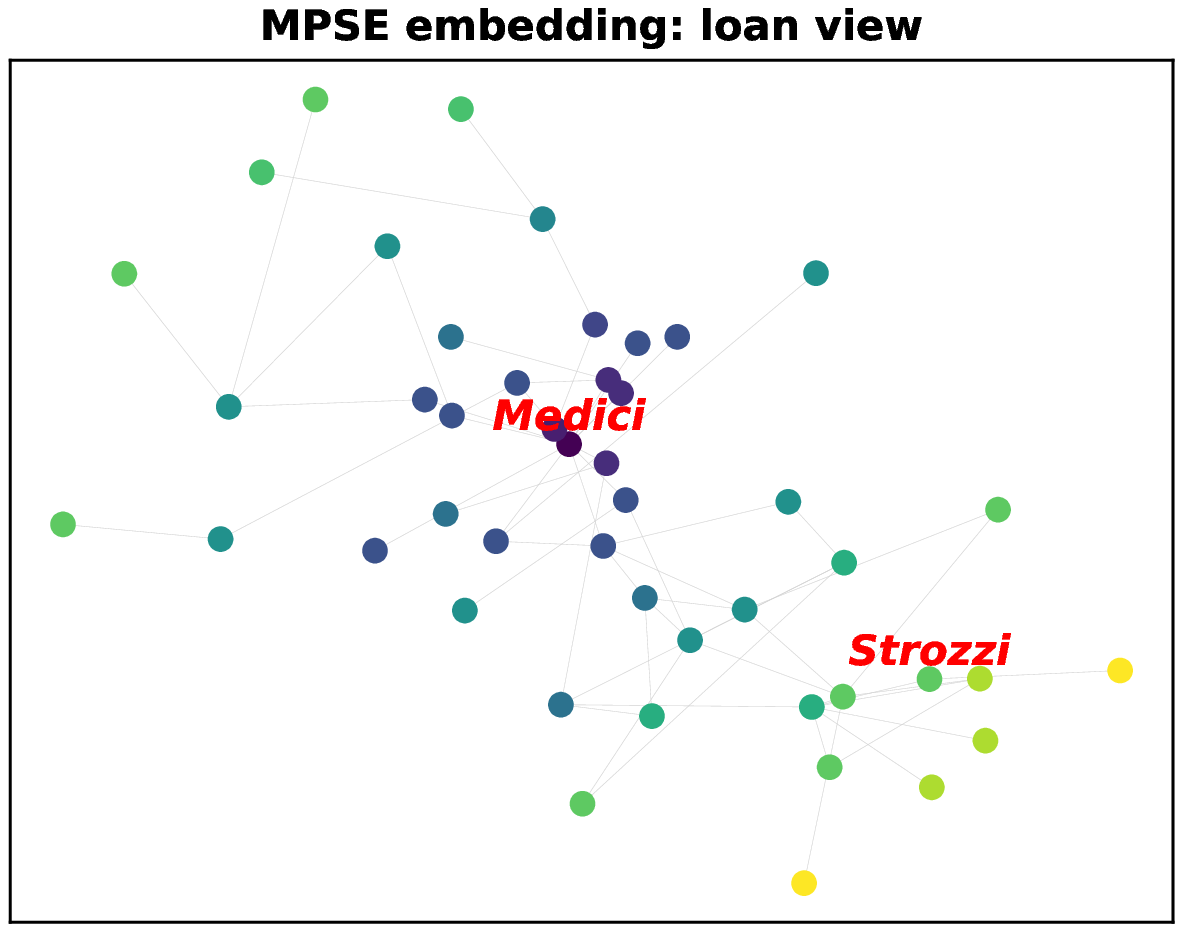}
\caption{MDS and MPSE embeddings of the  Florentine families dataset based on marriages and loans.
The first row shows the two 2D MDS embeddings (obtained independently of each other) and the second row  shows two projections of the 3D MPSE embedding from Fig.~\ref{fig:florence_3d}. 
The color indicates the true graph distance from the Medici family for each attribute with blue indicating close neighbors and yellow indicating distant ones (as described in the text). The MDS stress values are 0.70 for the marriage distances and 0.65 for the loan distances. The total MPSE stress is 0.78, with projection-wise stress values of 0.75 for marriage perspective and 0.81 for the loan perspective.}
\label{fig:florence_comparison}
\end{figure}

The first row of images in Fig.~\ref{fig:florence_comparison} shows the two 2D MDS embeddings
and the second row of images shows two projections of the 3D MPSE embedding. 
The first column corresponds to marriages and the second column corresponds to loans. 
The stress value for each embedding is the standard normalized MDS stress \eqref{eq:normalized-mds-stress}. Hence, the MDS and MPSE values are directly comparable to each other. 
Note that the individual normalized MDS stress for both perspectives in the MPSE embedding are close to the optimal normalized MDS stress values (computed individually for each attribute). This indicates that it is possible to visualize both attributes simultaneously with the MPSE method, without significantly increasing  stress values.

The colors in Fig.~\ref{fig:florence_comparison} indicate the true graph distance from the Medici family, with respect to marriage relations (the first column) 
and loan relations (the second column). 
A ``perfect'' embedding would result in dark blue colors near the Medici family and yellow for the nodes farthest away (with gradual blue-to-yellow transition in between). Note that we can obeserve this phenomenon both in the individual MDS embeddings and in the MPSE embeddings in  Fig.~\ref{fig:florence_comparison}. 
Different views of the 3D MPSE embedding are shown in Fig.~\ref{fig:florence_3d}, with the view corresponding to the optimal MPSE perspective of the marriage attribute on the left and the view corresponding to the optimal MPSE perspective of the loan attribute on the right. 



A bit of historical background can help us see some interesting results in our visualizations. The Strozzi family had been Florence's richest one, but ended up exiled and in ruin after the Medici took control of Florence's government in 1434.
The MPSE embedding, illustrated in Fig.~\ref{fig:florence_3d} and Fig.~\ref{fig:florence_comparison}, allows us a glimpse in this story. 
The Medici family occupies a central location in both the individual MDS and in the MPSE embeddings.
The Strozzi family occupies a similarly central position in the marriage perspectives, but not in the loan perspectives. This reflects the nature of their political power at the time, as both families has strong ties to other prominent families in Florence, but the Medici family was doing better financially by that time.
The better positioning of the Medici family is nicely illustrated in the MPSE 3D embedding in  Fig.~\ref{fig:florence_3d} in a way that cannot be illustrated by the MDS embeddings: the Medici family has a central position in the 3D embedding (with an average pairwise distance of 2.87), whereas the Strozzi family lies closer to the periphery (with an average pairwise distance of 4.05). Even though their position in the marriage network was firm, their position in the overall network was weaker.

\subsection{Credit Card Application Dataset}

Another real-world, high-dimensional dataset comes from anonymized credit card applications~\cite{ccdata}. The dataset contains information about customers of a bank (gender, education level, income, etc.). We randomly selected 1000 customers and focused on three parameters: annual income, gender, and level of education. The goal is to use MPSE to embed the dataset in 3D so that three different projections show us patterns based on the corresponding parameters. Since MPSE requires numeric input we modify some of the parameters a bit. For gender we map male and female to 0 and 1. For education level, we map lower secondary, secondary, incomplete higher and higher education to 0, 1, 2, 3. To avoid giving more importance to some of the features, each set of features is then normalized so that the scale of all MPSE perspectives is the same.

We compute an MPSE variable projections embedding (Algorithm~\ref{algo:varying}), 
after initialing it with  Algorithm~\ref{algo:initialization}, using the following parameters: maximum number of iterations $T = 200$, starting learning rate $\mu_0 = 1$, and stochastic constant $c = 0.004$. The results are shown in Fig.~\ref{fig:ccapp}. The first subfigure  shows one view of the resulting 3D embedding, and 
the remaining subfigures show the projections of the embedded dataset onto the gender, education level, and income views. 

The 3D embedding seems to capture well the different perspectives, shown in the 3 different projections. 
The gender view shows clearly separated clusters of male (orange) and female (blue color) customers. The education view groups customers with similar education levels (see the groups of squares, circles, and triangles).
The income view attempts to ``sort" the customers by income (high on the top, low on the bottom in this figure),
as captured in the change of color intensity from the top to the bottom. This view also shows us that education level and income are correlated (circles at the top), and that men have higher income than women (the orange clusters are higher than the blue clusters). 
Indeed the actual averages for male and female customers in this sample of size 1000 are $\$195000 $ and $\$152000$, respectively.

\section{Testing the scalability of Algorithms~\ref{algo:fixed} and \ref{algo:varying}}
In this section we test whether Algorithms~\ref{algo:fixed} and \ref{algo:varying} scale once the number of datapoints and the number of projections increase. For this purpose we create random datasets, that is we fix the number of datapoints $n$ and sample $n$ points uniformly from a solid ball with radius $1$ in 3D. Next, we fix the number number of projections $K$ and generate $K$ random projections $\mQ^1, \dots, \mQ^K$ of this data onto $\mathbb{R}^2$. From these projected datasets we generate the corresponding distance matrices $\mD^1, \dots, \mD^K$. The input for Algorithm~\ref{algo:fixed} is $\mD^1, \dots, \mD^K$ and $\mQ^1, \dots, \mQ^K$ and for Algorithm~\ref{algo:varying} the input is $\mD^1, \dots, \mD^K$. We consider the following 2 experiments for Algorithm~\ref{algo:fixed} and Algorithm~\ref{algo:varying}: The first experiment fixes the number of projections to be $K = 3$ and varies the number of datapoints $n$ between $100$ and $2000$ in increments of 100. We fix the max number of iterations $T = 100$ for the first experiment and compute the average time, and the number of times that the algorithm successfully finds the global minimum over 10 instances. 

\begin{figure}[h]
\includegraphics[width=0.49\linewidth]{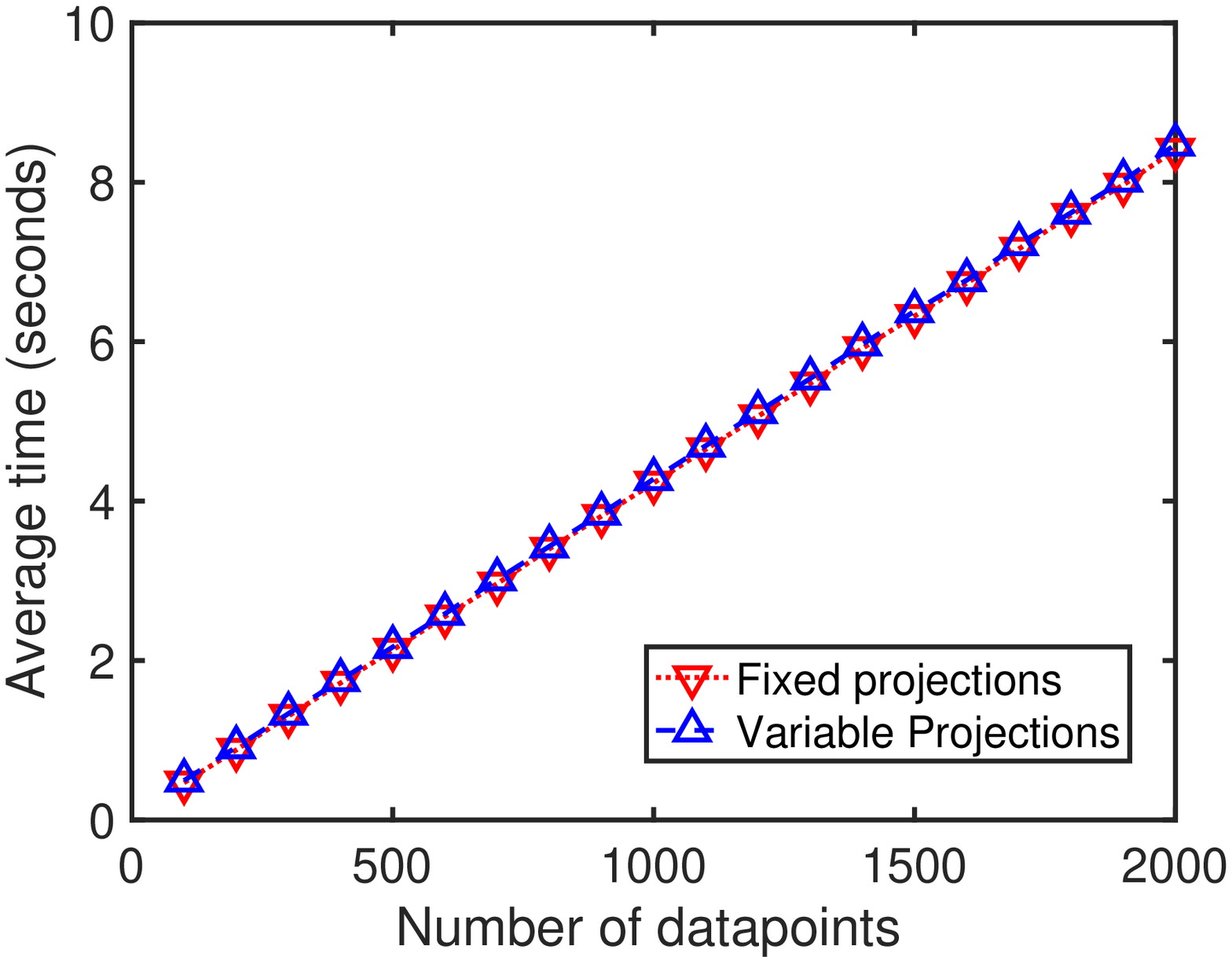}
\includegraphics[width=0.49\linewidth]{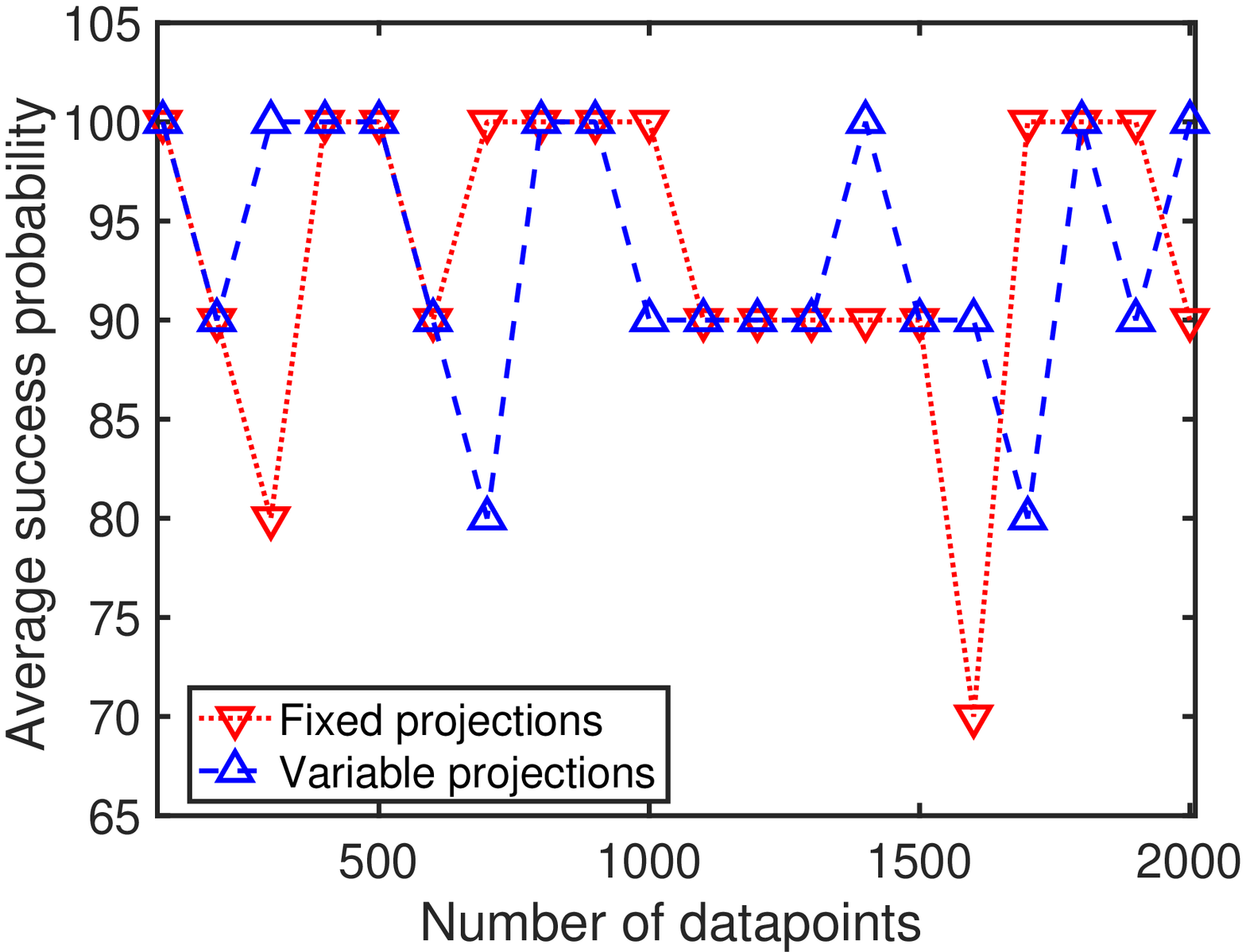}

\includegraphics[width=0.49\linewidth]{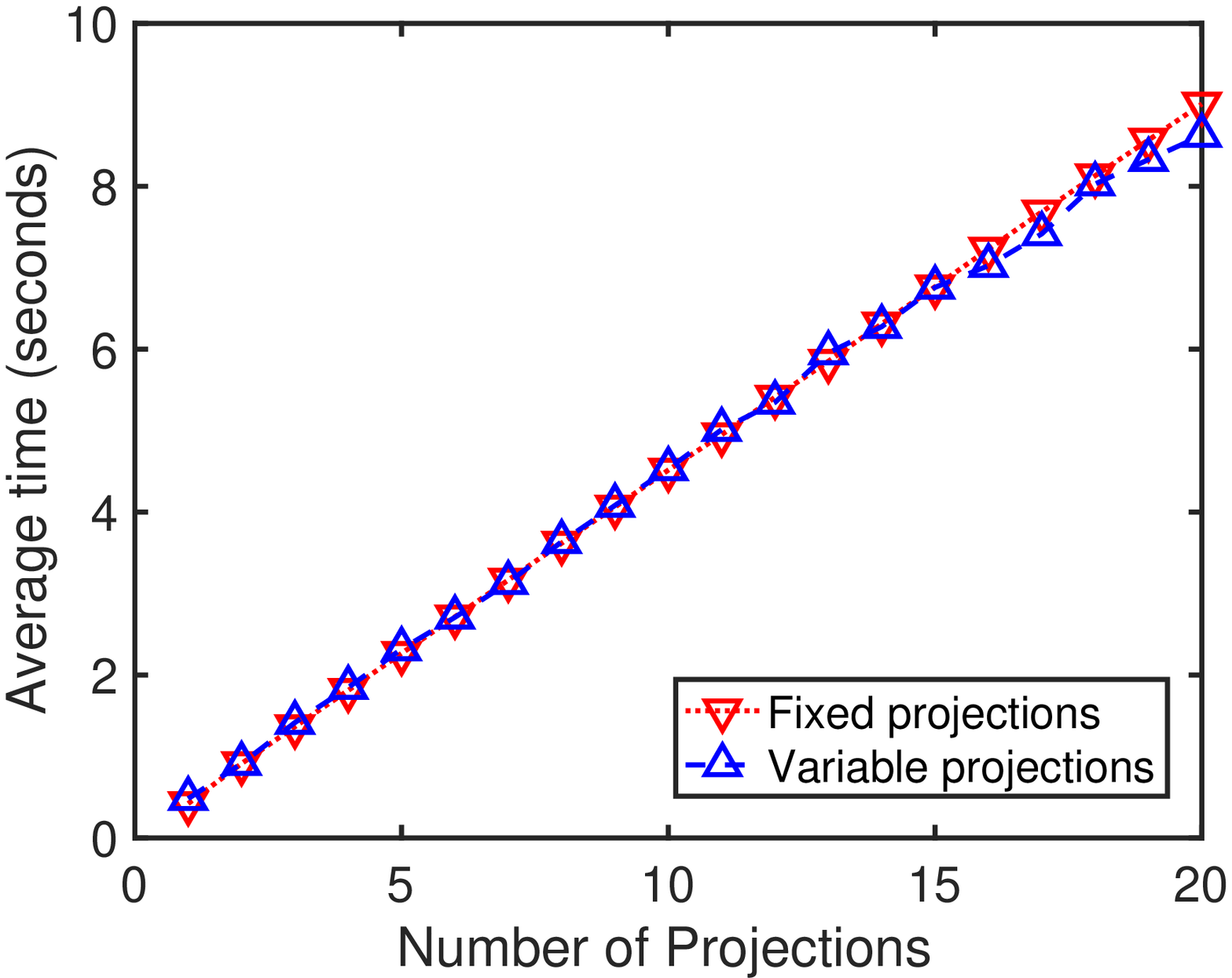}
\includegraphics[width=0.49\linewidth]{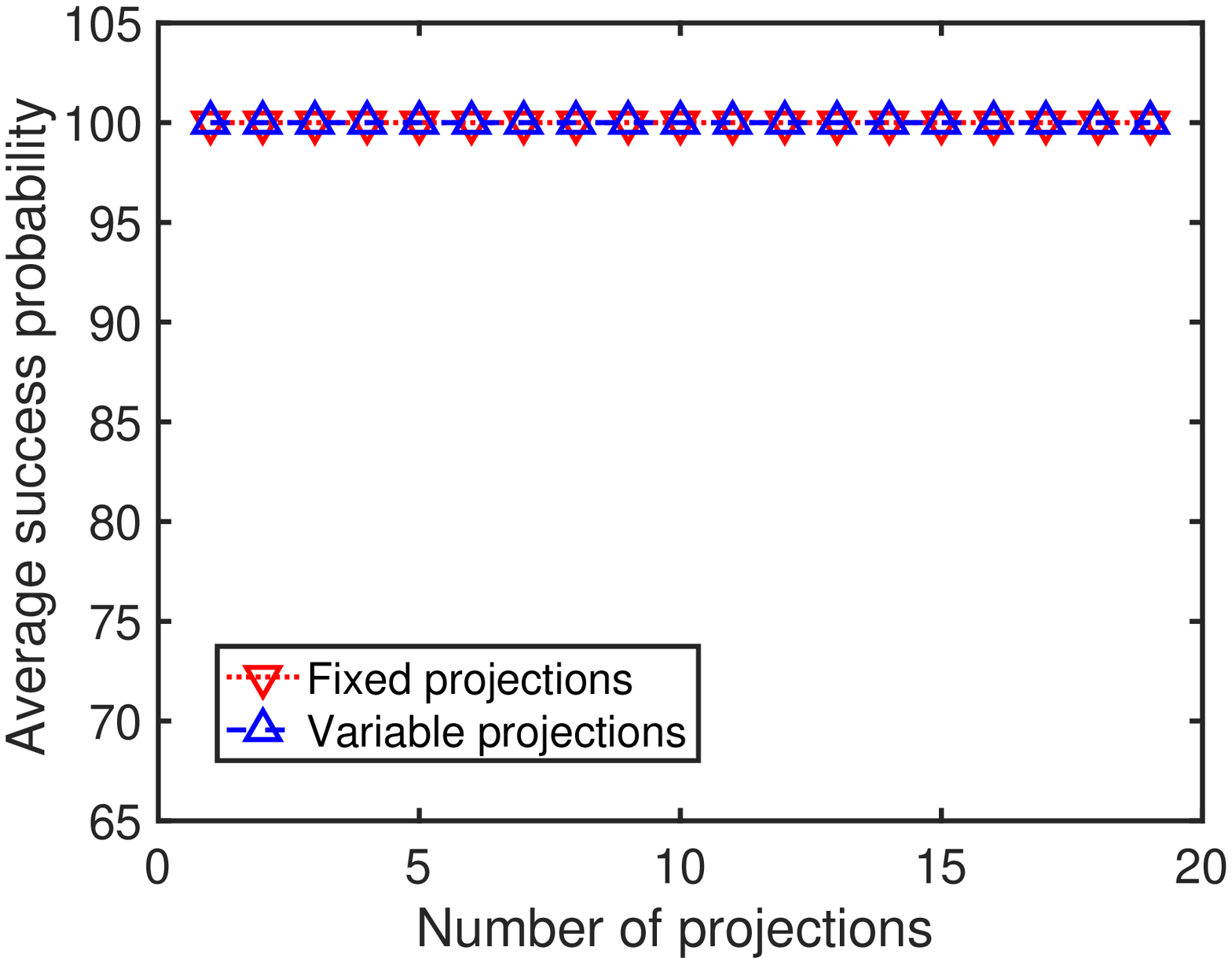}
\caption{Summary of the results for the scalability tests. The first row corresponds to the experiments where we fix the number of projections $K = 3$ and vary number of datapoints between $200$ to $2000$ with increments of $100$. The second row corresponds to the experiments where we fix the number of datapoints $n = 200$ and vary the number of projections $K$ from $1$ to $19$ with increments of $1$. The first column reports the average running time of the algorithms over 10 instances and the second row shows the success rates of the algorithms.}
\label{fig:scalability_test_1}
\end{figure}

The second experiment fixes $n = 200$ and varies the number of projections $K$ between $1$ and $20$ in increments of 1.
We fix the max number of iterations $T = 1000$ and compute the average time and number of times that the algorithm successfully finds the global minumum over 10 instances. The results are reported in Fig.~\ref{fig:scalability_test_1}. 
We remark that, as shown in Fig.~\ref{fig:scalability_test_1}, the running time of Algorithms~\ref{algo:fixed} and \ref{algo:varying} increases linearly with the increase of datapoints with Algorithms~\ref{algo:varying}.
Similarly, the running time of both algorithms increases linearly with the increase of the number of projections. The second column of Fig.~\ref{fig:scalability_test_1} shows that the increase in the number of datapoints and the increase of number of projections does not effect the success rate of both algorithms.

\begin{figure}[h]
\includegraphics[width=\linewidth]{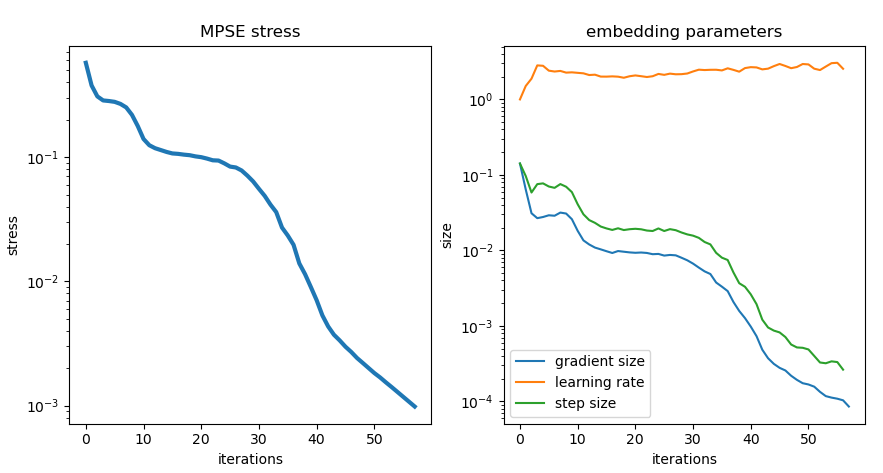}
\caption{Computation history of Algorithm~\ref{algo:fixed} for the ``1, 2, 3" data shown in Fig.~\ref{fig:teaser}. The left image shows the decay of total  multi-perspective MDS stress at each iteration. The right image shows the behavior of parameters (normalized gradient size, learning rate, and normalized step size) during the execution of the algorithm.
We used the 
following parameters for Algorithm~\ref{algo:fixed}: initial learning rate $\mu_0=1$,  stochastic constant $c=0.02$, and random initial embedding. The final MPSE stress of 0.001 was reached after 58 iterations, with projection-wise MDS stress values of 0.0009, 0.0008 and 0.0011.}
\label{fig:computation_history}
\end{figure}

Fig.~\ref{fig:computation_history} shows the computation history for  Algorithm~\ref{algo:fixed} when applied to the ``1, 2, 3" dataset which resulted in the images shown in Fig.~\ref{fig:teaser}. The left image shows multi-perspective MDS stress at each iteration. The right image shows the behavior of parameters (normalized gradient size, learning rate, and normalized step size) during the execution of the algorithm.

\section{\large Computation of the gradients}
\label{sec:relevant_math}

In this section, we present and derive the formulas for the gradient functions of the multi-perspective MDS stress function \eqref{eq:multi-perspective-mds-stress}. Our purpose is to assist the reader who wishes to implement these algorithms. Otherwise this section can be skipped.


For a given $n \times n$ dissimilarity matrix $\mD=[\mD_{ij}]_{i,j=1}^n$, the classical (metric) MDS stress function $\sigma^2:\mathbb{R}^{T \times n}\to\mathbb{R}$ is given by \eqref{eq:mds_stress}. Assume the stress function for $\mY = [y_1,y_2,\dots,y_n]\in \mathbb{R}^{T \times n}$ instead of $\mX$, we plan to use $\mX$ for the MDSE.

For a function $f:\mathbb{R}^{T \times n}\to\mathbb{R}$, we denote by $D_{y_i} f(Y) : \mathbb{R}^T\to\mathbb{R}$ the derivative operator of $y_i \mapsto f(Y)$ at $Y$.
Since
$$\|(y_i+\delta)-y_j\|^2 = \|y_i-y_j\|^2+2(y_i-y_j)^T\delta+\mathcal{O(\|\delta\|}^2),$$
it follows that
$$D_{y_i} \|y_i-y_j\|^2 = 2(y_i-y_j)^T$$
and
$$D_{y_i} \|y_i-y_j\| = D_{y_i} \sqrt{\|y_i-y_j\|^2} = \frac{(y_i-y_j)^T}{\|y_i-y_j\|},$$
so that
\begin{equation}
    D_{y_i} \sigma^2(\mY; \mD) = 2 \sum_{j\neq i} \frac{\| y_i - y_j \| - \mD_{ij}}{\|y_i-y_j\|} (y_i-y_j)^T.
\end{equation}
The gradient $\nabla_{y_i} \sigma^2(\mY; \mD) \in \mathbb{R}^T$ is given by
\begin{equation}
    \label{metric_mds_gradient}
    \nabla_{y_i} \sigma^2 (\mY; \mD) = 2 \sum_{j\neq i} \frac{\| y_i - y_j \| - \mD_{ij}}{\|y_i-y_j\|} (y_i-y_j).
\end{equation}

Suppose now that $Y=[P(x_1),P(x_2),\dots,P(x_n)]$, where $x_1,\dots,x_n\in\mathbb{R}^S$ and $P:\mathbb{R}^S\to\mathbb{R}^T$ is differentiable. We write $X=[x_1,x_2,\dots,x_n]$ and $Y=Y(X,P)$. We have
$$D_{x_i} \sigma^2(Y(X,P);D) = D_{y_i}\sigma^2(Y;D) DP(x_i),$$
where $DP(x)\in\mathcal{L}(\mathbb{R}^S,\mathbb{R}^T)$ is the derivative of $\mP$ at $x\in\mathbb{R}^S$. It follows that
\begin{equation}
    \nabla_{x_i} \sigma^2(Y(X,P);D) = \nabla P(x_i) \nabla_{y_i}\sigma^2(Y;D),
\end{equation}
where $\nabla P(x) = [\nabla P_1(x),\dots,\nabla P_T(x)]\in \mathbb{R}^{S\times T}$.

Suppose now $K$ distinct $n\times n$ dissimilarity matrices $\mD^{(1)}, \mD^{(2)},\dots, \mD^{(K)}$ are given for the same $n$ objects. We write $\cD=\{\mD^{(1)}, \mD^{(2)},\dots, \mD^{(K)}\}$. We wish to explore how each dissimilarity matrix $D^{(k)}$ can arise from a perspective mapping $P^{(k)}:\mathbb{R}^p\to\mathbb{R}^q$. In our first algorithm, the perspective mappings are known, and the only requirement is that each mapping $P^{(k)}$ is differentiable. In the second algorithm, the mappings are no longer known, but are assumed to be parameterized.

The multi-perspective MDS stress function $S:\mathbb{R}^{S \times n}\to\mathbb{R}$ is given by
\begin{equation}
    \label{multi-perspective_MDS_stress}
    S(\mX; \cP,\cD) = \sum_{k=1}^K \sigma^2(\mY(\mX,\mP^{(k)});\mD^{(k)}).
\end{equation}
so that
$$ D_{x_i} S(\mX, \mP, \cD) = \sum_{k=1}^K D_{y_i} \sigma^2(\mY;\mD^{(k)}) D \mP^{(k)}(x_i)$$
and therefore
\begin{equation}
    \label{X_gradient_general}
    \nabla_{x_i} S(\mX;\mP,\cD) = \sum_{k=1}^K \nabla \mP^{(k)}(x_i) \nabla_{y_i} \sigma^2(\mY(\mX;\mP^{(k)});\mD^{(k)}).
\end{equation}

If the perspective mappings are linear, so that $P^{(k)}(x)=Q^{(k)}x $ for some $T \times S$ matrix $Q^{(k)}$, then $\mY(\mX,P^{(k)})=Q^{(k)} \mX$ and $\nabla P^{(k)}(x) = Q^{(k)}$ for any $x$. The gradients of $S$ reduce to
\begin{equation}
    \label{X_gradient_linear}
    \nabla_{x_i} S(\mX; \cP,\cD) = \sum_{k=1}^K Q^{(k)} \nabla_{y_i} \sigma^2(Q^{(k)} \mX; \mD^{(k)})
\end{equation}

In our second approach to optimizing \eqref{eq:multi-perspective-mds-stress}, it is not assumed that the perspective mappings are known, but instead that these belong to some parametric family of $C^1(\mathbb{R}^S,\mathbb{R}^T)$ maps. We assume that the set of allowed parameters $\mathcal{Q}\subset\mathbb{R}^r$ is open in $\mathbb{R}^r$. For each $q\in\mathcal{Q}$, we write $P^{(q)}:\mathbb{R}^S\to\mathbb{R}^T$. Furthermore, the mappings $P^{(q)}$ are assumed to be differentiable with respect to $q$, in the sense that the function $q \mapsto P^{(q)}(x)$ is differentiable for all $x\in\mathbb{R}^S$.

For a given $x\in\mathbb{R}^{S}$ and $q\in\mathcal{Q}$, we write $D_{q} P^{(q)}(x) \in \mathcal{L}(\mathbb{R}^S,\mathbb{R}^T)$ as the derivative of $q \mapsto P^{(q)}(x)$. 
Consider the function $q \mapsto \sigma^2(\mY(\mX,\mP^{(q)});\mD)$. The derivative is
$$ D_{q} \sigma^2(Y(\mX, \mP^{(q)}); \mD) = \sum_{i=1}^n D_{y_i} \sigma^2(\mY(\mX, \mP^{(q)});D) \mD^{q} \mP^{(q)}(x_i)$$
and its gradient is
\begin{equation}
    \nabla_q \sigma^2(\mY(\mX,\mP^{(q)});\mD) = \sum_{i=1}^n \nabla_q P^{(q)}(x_i) \nabla_{y_i} \sigma^2(\mY(\mX,\mP^{(q)});\mD)
\end{equation}

For linear correspondences $q \mapsto P^{(q)}(x)$, $D_q P^{(q)}(x) \in \mathcal{L}(\mathbb{R}^r,\mathbb{R}^T)$ is given by $D_q P^{(\cdot)}(x) \mapsto P^{(\cdot)}(x)$ and therefore
$$\nabla_q \mP^{(q)}(x) =[\mP^{(e_1)}(x),\dots, \mP^{(e_r)}(x)]^T$$ 
where $e_1,\dots,e_r$ is the standard basis for $\mathbb{R}^r$.

For $r=ST$, for a given $q\in\mathbb{R}^r$, we rewrite $q$ as a $T$ by $S$ matrix $\mQ$, so that $P^{(q)}(x) = \mQ x$. we have
\begin{equation}
    \nabla_Q \sigma^2(\mY(\mX,\mP^{(q)});\mD) = \sum_{i=1}^n \nabla_{y_i} \sigma^2(\mQ \mX;\mD) x_i^T
\end{equation}
This can be rewritten as
\begin{equation}
    \nabla_{\mQ} \sigma^2(\mY(\mX,\mP^{(q)});\mD) = \nabla_Y \sigma^2(\mQ\mX;\mD)\mX^T
\end{equation}
For stochastic gradient descent schemes, an approximation to \ref{metric_mds_gradient} can be constructed as follows: the $n$ objects are divided into batches; if object $i$ belongs to batch $b(i)$ with size $|b(i)|$, then the approximate gradient is given by
\begin{equation}
    \label{metric_mds_gradient_approximate}
    [\tilde{\nabla} \sigma^2 (\mY; \mD)]_i = 2 \frac{n}{|b(i)|}  \sum_{j \in b(i),\ j\neq i} \frac{\| y_i - y_j \| - \mD_{ij}}{\|y_i-y_j\|} (y_i-y_j).
\end{equation}


\section{Conclusions and Future Work}
We described a generalization of MDS which can be used to simultaneously optimize multiple distance functions defined on the same set of objects. The result is an embedding in 3D space with a set of given or computed projections that show the different views. This approach has applications for visualizing abstract data and  multivariate networks. 
It would be interesting to experiment and test whether MPSE can compete with current techniques for 3D reconstruction from 2D images. For example, one could combine the MPSE approach with a multi-way matching algorithm  \cite{huroyan2018mathematical,pachauri2013solving} and a feature extraction algorithm~\cite{lowe1999object} to automatically detect points of interests in each image, match them, and place them in 3D to recover the 3D figure.
The MPSE webpage \url{http://mpse.arl.arizona.edu} provides 3D visualizations with interactive examples, and a functional MPSE prototype; 
 \url{https://github.com/enggiqbal/MPSE} provides the source code.  

\section*{Acknowledgements}

This  research was supported in part by  National Science Foundation grants CCF–1740858 and DMS-1839274.

We thank the participants in the ``Visualizing Dynamic Graphs in VR" working group at Shonan Seminar 131 ``Immersive Analytics for Network and Trail Sets Data Analysis": David Auber, Peter Eades, Seokhee Hong, Hiroshi Hosobe, Stephen Kobourov, Kwan-Liu Ma, and Ken Wakita. 
%


\bibliographystyle{abbrv-doi}
\bibliography{multiview}

\end{document}